**ORIGINAL ARTICLE**

# Daytime Photometry of Starlink Satellites with the Huntsman Telescope Pathfinder

**Sarah E. Caddy**[1,2,3]  · **Lee R. Spitler**[3,4]

**Abstract**

The rapid increase in satellite launches in recent years, and the pressure of launches planned into the next decade, demands an improvement in the efficiency of space domain awareness facilities. Optical facilities form an important component of global space domain awareness capabilities, however traditional optical telescopes are restricted to observing satellites during a small twilight window. In this work we explore expanding this operational period to encompass the entire day to dramatically improve the observing opportunities at a single site. We explore daytime space domain awareness observations with the Huntsman Telescope Pathfinder, an instrument built using predominantly off the self components, and Canon telephoto lenses. We report successful detections and photometric light curves of 81 Starlink satellites from Sun altitudes ranging 20° to midday. Starlink satellites are found to be particularly bright at $3.6 \pm 0.05$ mag, $\sigma = 0.6 \pm 0.05$ mag in Sloan r′, or $\sim 11\times$ brighter than twilight conditions. We conclude this surprising observed brightness is due to the contribution of Earthshine beneath the orbiting satellites. We also compare our observations to existing satellite optical brightness models and find that satellite optical brightness during the day can only be well described by a model including an Earthshine component. We find that observed light curves are more complex than simple geometric models predict, but generally agree within an order of magnitude. Finally we suggest improvements to satellite optical brightness models by incorporating weather data to measure the actual Earthshine under a satellite.

# 1 Introduction

Space is getting crowded. With thousands of satellites planned for launch over the next decade, the rise of satellite mega constellations, and an increase of 6 times the objects in Low Earth Orbit (LEO) in recent years alone[1] (see data from United Nations Office for Outer Space Affairs Fig. 1), we must improve our capabilities to monitor space assets [1–3]. This capability is known as space domain awareness (SDA).

The biggest contributor to recent LEO satellite launches are Starlink satellites. Starlink is the worlds largest satellite constellation[2] and make up a significant proportion of objects in orbit—as of April 2024, over half of all active satellites belonging to Starlink alone.[3] This can be placed in context when considering that Starlink satellites, as of 2024, have only been launched for just over 7% of the entire time humankind has been launching satellites since Sputnik 1 in 1957.

Improvements to the global capacity to monitor satellites in LEO, including the increasing population of Starlink satellites, will be vital to ensure a safe and sustainable future in Space for generations to come [2]. The majority of SDA capabilities currently fall into two broad sensor categories. Radar, and optical observations [3, 4]. Radar technologies include both passive and active radar [3]. Different types of active radar facilities have a broad range of capabilities. While they do have poorer angular resolution than optical facilities to determine a targets location on sky, ranging may be used to determine the distance to an object as well. If Doppler shift information is collected the objects, radial velocity can also be determined [4, 5]. In some instances the objects size and attitude can be determine from the radar cross section [3]. In addition, some phased arrays are capable of tracking multiple objects at a time [3, 5].

Globally, the uptake of radar facilities for SDA has been less than optical facilities. This is largely due to trade-off between cost and performance [4]. Radar facilities are expensive to build and maintain. Their capabilities vary greatly depending on the design, they are often optimised for the tracking of specific targets [3], and they are limited in terms of orbital range (most facilities can only operate up to the edge of LEO at 2000 km[4]) As a result optical SDA facilities have become a cornerstone of global SDA capabilities, widely used by both governments and multinational collaborations, to individuals and private industry [4]. This has also been driven in part by the rise of commercial off the shelf technologies (COTS) driving down prices and development times [4].

Optical SDA observations however, have their own limitations. Historically, optical observations of LEO satellites from the ground have been restricted to terminator illuminated conditions [6–8] or "twilight". This is a configuration where the observer is in the Earth's shadow just after Sunset, and the satellite is illuminated by

---

the Sun. Further restrictions arise where LEO satellites pass into the Earth's shadow and become too faint for an optical SDA facility to detect them via reflected Sunlight [7].

In addition, when restricted to twilight, some satellites may only be visible for a few minutes –if at all–every few weeks from a given observing location [9]. To illustrate the restrictions that this imposes, one can consider a sample of LEO satellites and calculate observing opportunities per minute both during the day and in twilight. Using the sample of 755 Starlink V1.5 satellites[5] observable in a typical 24 h period in summer at Macquarie University Observatory in Sydney, Australia, 74% of observable passes occur during the day, and only 26% are twilight illuminated passes.

If SDA facilities could operate effectively all day instead of just during twilight conditions, this would greatly improve the productivity of global optical SDA facilities. Daytime observation would allow us to take advantage of all Sunlit passes of satellites above an observing location (both during twilight and during the day). More frequent observations of satellites combined with GPS timing enabled by daytime observations will ultimately result in a decrease in the uncertainty of satellite orbital elements, and an accurate orbital determination [7].

The significance of this capability is emphasised by the continued monitoring of debris from the collision of a decommissioned Russian military satellite and an Iridium communications satellite in 2009. This collision subsequently resulting in a dangerous cloud of debris in LEO. The significance of this single event reflects the importance of maintaining accurate orbital determinations, and reducing the risk of in orbit collisions [10].

In addition to a single detection of a satellite in LEO, tracking and producing accurate photometric light curves provides another dimension of information that is useful for SDA analysis. Many studies conducted have demonstrated the importance of light curves as an essential tool for characterising and classifying LEO satellites during twilight and Geosynchronous Orbit (GEO) satellites at night based on their material properties, as well as capturing both position and dynamics information [11–23]

The potential for backwards propagating the complex light curves of LEO satellites to understand their orientation without the need for resolving the target is one such use case of photometric light curves [11, 14]. Self shadowing on the surface of the satellite as well as differences in the scattering properties of different materials on the satellites surface may lead to complex divots and peaks in the light curve which can be used as signatures to match the light curve to rendered models of the satellite in orbit [14, 24]. This is particularly useful for satellites, rocket bodies and debris which no longer have control of their attitude and deemed space junk. If these objects are out of control and de-orbiting, accurate information about the objects position, attitude and velocity during descent will aid in predicting the point of impact [14]. In addition to LEO objects, there have also been multiple works studying the use of photometric light curves of GEO

---

[5] https://github.com/Forrest-Fankhauser/satellite-optical-brightness/blob/main/data/brightness_config_list.csv

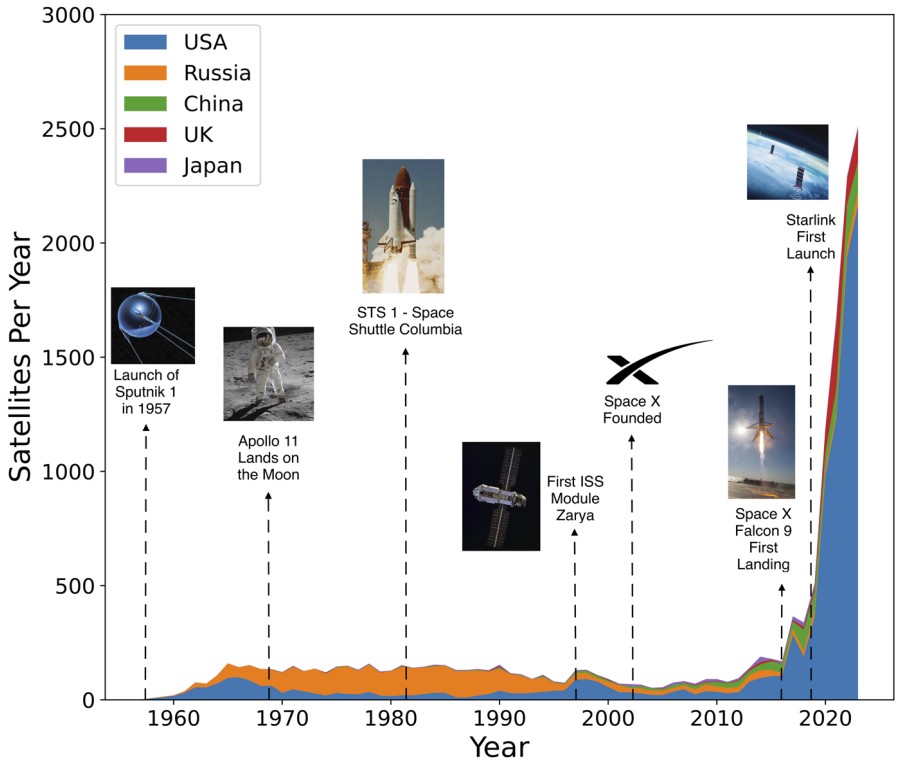

**Fig. 1** The number of satellites launched per year by the countries who are the 5 largest contributors: The USA, Russia, China, the UK, and Japan. All other countries make up < 1% of the total world launch rate. Starlink satellites now make up over half the total number of active satellites in orbit. Data spans 1957 with the launch of Sputnik 1 to the total for 2023. The launch of Starlink satellites now contributes to more than 2500 launches per year, a rate higher than any other year in history. Data Source: United Nations Office for Outer Space Affairs (2024). Image Credits (left to right)—Physics World, NASA, Space X, Getty Images

objects for classification [12, 15, 18, 19]. Observing GEO satellites during the day is out of the scope of this work, but will be the focus of future exploration.

Light curves of LEO and GEO satellites have also been studied during twilight and night respectively to determine their rotation about their center of mass, particularly for satellites that are tumbling out of control [16, 20–23]. Almost all of these satellites in LEO and GEO are too small to directly resolve, and so light curves remain the most effective way of monitoring the rotational period. Satellites without active stabilisation begin to tumble under the influence of forces that are poorly understood which in combination with flat reflective surfaces on the body of the satellite produces periodic glints which can be used to determine rotational velocity [18, 25, 26].

Daytime observations are expected to be boosted in brightness by Earthshine, following the results of [24]. Earthshine consists of two components: (1) re-radiated infrared light and 2) scattered Sunlight. At ∼ 0.3−2 μm Earthshine is dominated by reflected Sunlight and predominantly bluer in colour [27]. Earthshine has been shown through simulations and comparison to data in twilight terminator illuminated conditions to contribute to satellite optical brightness in LEO [24]. In addition, in conditions where Earthshine is brightest, typically where there is ice or large systems of clouds below the satellite in LEO, [6] predicts that satellites may be illuminated comparably to terminator illuminated conditions.

Optical daytime observations however, are challenging. The bright sky background and increased atmospheric turbulence due to large temperature gradients during the day result in low signal-to-noise (SNR) detections, and bright detection limits [6, 7, 9, 28–30]. For optical daytime observations to be successful, a well calibrated robotic telescope mount is required, equipped with photometric broadband filters to reduce the bright sky background [28]. Further, recent improvements in Complimentary Metal Oxide Semiconductor (CMOS) detectors has resulted in cameras that now rival the quality of Charged Couple Device (CCD) detectors [31]. They have enabled the high frame rate observations required to take multiple rapid observations of a moving target, preventing over exposure and enabling stacking of data to achieve higher SNR detections [28].

So far, detecting a LEO satellite during the day has been successfully attempted by a number of groups [7, 9, 29, 32]. These attempts have been single detections, and none have produced light curves (with the exception of Caddy et al. (2024) [28], the authors catalyst for this work). These investigations span the optical and near infrared - most were motivated by the need to 'close the daylight gap' [7] and build cathermal SDA facilities that operate both during twilight, night (for GEO) and during the day (e.g., [29, 33]).

Rork et al. (1982) [32] reports an attempt in 1982 using an ETS 31-inch f/5 telescope with an RCA Ultricon silicon vidicon television camera. They achieve a detection limit of 8.3 magnitudes in visual wavelengths and track 20 large bright satellites successfully through Sun angle ranges between 21 to 142 ° and magnitudes of 1 to 8.9 at visual wavelength. Sun altitudes for observations are not recorded, but observations are describes as having occurred between "full daylight" and Sunset.

Estell et al. (2019) [7] reports observations of bright LEO targets at low Sun altitudes up to 14 °. They make use of both a 0.4m DFM Ritchey Chreien telescope at the University of Michigan's Angell Hall Observatory and a Canon T3i DSLR and with no filter in a non-tracking mode to detect the Chinese space station. In addition, they test a Takahashi FSQ-85EDX 85 mm aperture f/5.3 telescope, a ZWO ASI1600MM Pro camera and a Sloan r' filter. Only 8 passes are attempted, with 7 detections successful of objects in the "Brightest 100" Celestrak category.

Zimmer et al. (2020, 2021) [6, 9] describes the surprising detection of Starlink satellites as bright as 2.6 magnitudes in an 570nm long pass filter with a 12.5" f/7 Planewave CDK telescope and a ASI183MM Pro camera as well as a selection of long pass filters.

At short wave infrared wavelengths where the daytime sky is fainter there have also been successful detections. Shaddix et al. (2021) [29] using the half-meter "Aquila" cathemeral telescope system successfully reports observing GEO and LEO objects at Solar noon, down to a limiting magnitude of 11 in equivalent visible wavelengths. Of these observations, LEO satellite detections during the day have either been limited to low Sun altitudes [7], near infrared wavelengths [29, 33], and single images [9]. Thus far, to the authors knowledge at the time if writing, there have been no published photometric light curves of daytime SSA observations either in the optical or near infrared wavelengths.

In this work, we describe an optical SDA facility, The Huntsman Telescope Pathfinder [28] (see Fig. 2), which we use to demonstrate the successful detection and collection of Starlink light curves during the day, at a range of Sun altitudes, including midday. The Huntsman Telescope Pathfinder is a test facility located at Macquarie University, Sydney Australia. It is designed to test experimental science cases for the Huntsman Telescope, a robotic < 0.5 m class research facility located at Siding Spring Observatory, Australia on Gamilarray, Wiradjuri and Wayilwan country [34]. While the Huntsman Telescope was originally designed for low surface brightness imaging at night, its high étendu, large field of view and good stray light control are properties that are ideal for daytime SDA.

In the work of Caddy et al. (2024) [28] we show that the Huntsman Telescope Pathfinder is capable of 1–10% photometry during the day, so extending this work to capturing light curves of satellites is a natural next step. We detail our observing methods, and present light curves for 81 Starlink satellites spanning version V1.0 to V2.0, and explore the factors influencing the brightness of satellites during the day, and the implications of this work in the era of satellite mega constellations.

## 2 Method

### 2.1 Hardware, Software and Observing Technique

Observations are conducted over January to March of 2024 using the Huntsman Telescope Pathfinder at Macquarie University Observatory, Sydney, Australia. This instrument is identical to the Huntsman Telescope, except it comprises of only one Canon lens, as opposed to an array of 10. We use a single Canon 400 mm f/2.8 lenses with a field of view (FOV) of $1.89° \times 1.26°$ and a pixel scale of 1.24". Following the results of Caddy et al. [28], the lens has an filterwheel containing a Sloan r' filter which we use for this work in order to minimise the sky brightness during the day. We use an Astromechanics focuser,[6] and a ZWO ASI183MM Pro[7] camera.

In order to ensure the target is near the center of the FOV, it is essential that the pointing of the mount, the location settings, the computer time, and the Two Line Element sets (TLEs) are as accurate as possible. Following some trial and error and the generous assistance of the team at Software Bisque, we achieved our first reliable tracking of the International Space Station (ISS). We use an MEII mount from Software Bisque, and TheSkyX for mount control. Using the TPoint add on, we ensure an accurate polar alignment, and conduct a 400 point pointing model at night to reduce the pointing RMS to 30" after running the command "supermodel"

---

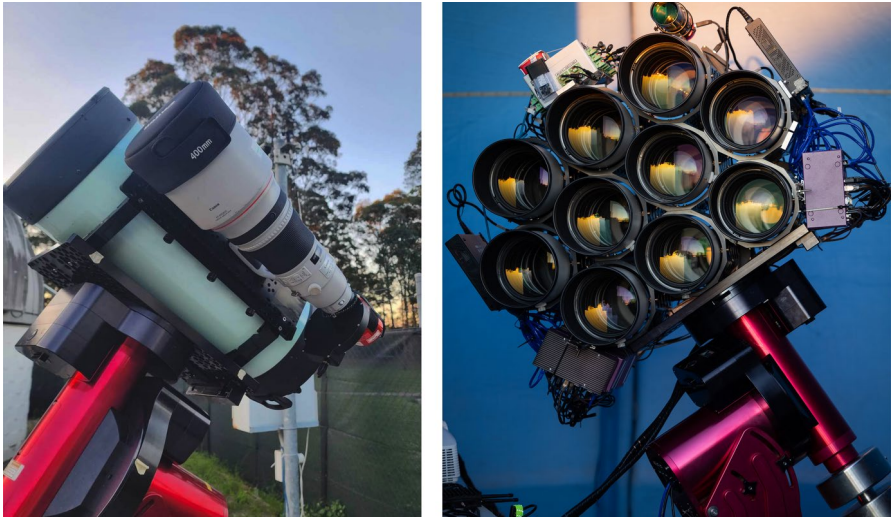

**Fig. 2** Left, The Huntsman Telescope Pathfinder consisting of a single Canon lens used in this work. The Rowe–Ackermann Schmidt Astrograph is not used in this work. The unit is located at Macquarie University Observatory, Sydney, Australia. Right, The Huntsman Telescope facility located at Siding Spring Observatory, Australia. The telescope consists of 10 Canon 400 mm f/2.8 lenses which are configured to cover the same field of view of 1.89° × 1.26° with a pixel scale of 1.24'

which calculates and applies higher order corrections based on the characteristics of the mount. We set up a NTP server to ensure the computer system time is accurate to within ~1ms. We use the NearMap[8] program to determine the precise longitude and latitude of the mount, and we determine an estimate for the observing location altitude using a topographic map.[9] TLE's are downloaded from Celestrak[10] every hour to ensure an accurate ephemeris is known for each target. In addition, we make use of SkyTrack[11] to calculate and filter for the satellite passes that avoid obstructions on our local horizon.

For all observations, active tracking is used where by the target is approximately centered within the FOV and remains stationary in the image over the duration of the pass. Early attempts using the leap frog method to reduce the strain on the mount were unsuccessful. In the leap frog method, the telescope is driven to the path in front of the satellite, and the mount remains stationary as the satellite transits through the FOV. Due to the large number of false positives dubbed as 'angels' [32] originating from seed packets, leaves, insects and ice crystals, it was found to be impractical to attempt to distinguish between a satellite and a false positive detection in leap frog mode.

---

[8]  https://www.nearmap.com/

[9]  https://en-au.topographic-map.com/

[10]  https://celestrak.org/

[11]  https://www.heavenscape.com/

In this work we read out the full frame of the detector in a $5496 \times 3672$ array at typically 2–3 frames per second set by the speed to write to a networked disk directly. The exposure time is tuned manually for each target to ensure the sky background is not over exposed throughout the duration of the pass, and is typically between $\sim 0.01 - 0.001$ seconds in r' band depending on the Sun separation angle (the angular distance between the Sun and the target on sky). The focus is manually fine tuned for every observation using a reference star of approximately similar magnitude to the observed target, as temperature fluctuations during the day result in large, continuous focus offsets [28]. The NORAD ID is recorded for every object, and the time of observation is recorded in the FITS header to ensure that the altitude and azimuth of the pass can be calculated for post processing.

## 2.2 Data Reduction

Photometric calibration flat field data are taken at twilight after each observing session. Flats are taken during twilight to reduce the prominence of the sky gradient in the flats taken during the day [28]. Approximately 100 Sloan r' flats are median combined to create a master flat that is used to calibrate the science frames.

For this work where only $\sim 100 - 200$ frames are collected per target, we do not design an algorithm for false positive detection. Instead, a program was written in python that allows a user to quickly and dynamically select the target on screen, and automatically save the pixel location of the click to a json file. The json file can then be loaded by a data reduction script which applies the calibration files to each science frame, creates a $100 \times 100$ pixel postage stamp of each target about the click locations, and fits a 2D Gaussian profile, which yields a centroid on the target for the purposes of aperture photometry. Following the results of [28] we use a fixed aperture radius of 11 pixels for all images to maximise the target flux in the aperture over a range of seeing conditions (our targets range from $\sim 3$ to 6 arcseconds FWHM typically). This overly generous aperture reduces impact from variable seeing, which otherwise would require variable aperture corrections to be applied. The calibrated apparent magnitude is then:

$$r' = r'_{\text{inst}} - ZP_{r'} - k_{r'}(X)$$

where $r'_{\text{inst}}$ is the instrument magnitude in units of $e^-/s$, r' is the calibrated magnitude, $ZP_{r'}$ is the system zeropoint, $k_{r'}$ is atmospheric extinction coefficients, and $X$ is the airmass at the time of the measurement. As we do not have multi-wavelength observations of our targets, we do not perform a colour correction.

A detailed exploration of the Huntsman Pathfinder's photometric accuracy during the day can be found in [28]. For Sloan r' we adopt the measurements of [28] of $ZP_{r'} = -21.7$ mag and $k_{r'} = 0.34$. The total photometric accuracy is reported to be a median of 0.05 mag in Sloan r' with a standard deviation of 0.03 mag throughout the duration of the 7 month campaign described in [28]. There are however, some caveats to this photometric accuracy applied to this work. We do not take separate reference star calibration observations, and instead use the median zeropoint, extinction coefficient, and colour coefficient over the [28] work. We note that the zeropoint calibration is out of date in the sense that e.g., dust accumulating on the lens since the calibration campaign will lead to a zeropoint error most likely in the sense that the reported apparent magnitudes in this work are fainter than reality. We also note that the median extinction coefficient is high for our system, likely due to the atmospheric conditions as the site is on a coastal, urban location. We note the caveat that this is not an ideal location for photometric observations, as it is likely that changing aerosol content in the atmosphere will cause variation in the extinction over the course of the day. However, we present this work as a proof of concept, and further observations will be conducted at a better site at Siding Spring Observatory for further work, with photometric zeropoints derived for each day of observations.

During processing, for each target observation the TLE is retrieved by special data request in Celestrak for the time and date when an image is taken. The TLE, observatory site location details and timestamps for each frame are input into the `skyfield` [35] EarthSatellite and Topos class instances, and this is used to derive the satellite altitude in meters at the time of observation, the longitude and latitude of point on the surface of the Earth directly below the satellite and the satellite altitude and azimuth from the observers location at the time of observation. `Astropy` [36] is used to compute the Sun coordinates with respect to the observer, from which we calculate the angular separation of the target from the Sun on sky.

## 2.3 Satellite Optical Brightness Modelling

We compare our observations against an existing satellite optical brightness model called `lumos-sat`[12] described in [24]. The software is open source, and critically, includes Earthshine estimates for the calculated brightness of Starlink satellites. The software was not designed to produce daytime estimates of satellite brightness, and so minor alterations are made to remove exceptions thrown when the Sun altitude of an observation is above 0°. We note a caveat of this model in that it does not include Rayleigh scattering in the atmosphere, which is a possible direction for future work. The geometry of the problem implemented in `lumos-sat` remains the same for the day and is not changed. The code models satellites as a collection of opaque surfaces, with a Bidirectional Reflectance Distribution Function (BRDF) [37] describing the angular distribution of light reflected from surface and takes the general form:

$$BRDF = f_r(\hat{w}_i, \hat{w}_o) \equiv \frac{1}{\cos(\phi_o)\cos(\phi_i)L_i}\frac{\partial L_o}{\partial \hat{w}_i}$$

where $L_i$ is the ingoing radiance and $L_o$ is the outgoing radiance. $\phi_i$ is the angle between the surface normal and the vector to the source $\hat{w}_i$, and similarly $\hat{w}_o$ is the angle from the surface normal to the observer. The BRDF function describes the ratio of the spatially distributed, reflected radiance outgoing from a surface, to the incident irradiance. For example a shiny metallic surface or body of water may exhibit specular reflections at particular angles, compared to diffusely reflected light

---

from a forest canopy, sandy beach, or coarse material on a satellite body. We consider two BRDF's in this work, as described in [24]. The Phong model is utilised to represent light scattered from the surface of the Earth and combines a lambertian BRDF where light is scattered equally or diffusely in all directions ($\frac{K_d}{\pi}$) with a specular peak component defined by:

$$BRDF = \frac{K_d}{\pi} + K_s \frac{n+2}{2\pi} (\hat{w}_r \cdot \hat{w}_o)^n$$

where the $K_d$ parameter controls of the magnitude of the diffuse component, $K_s$ the magnitude of the specular component, and n controlling the width of the specular peak. $\hat{w}_r$ and $\hat{w}_o$ are the reflected and outgoing unit vectors respectively. Based on the recommendations of [24] a more complex Binomial BRDF model is used for satellite surfaces. The model is fit to data provided by SpaceX which was gathered in a lab specifically for the V1.5 satellite and is derived in [24] and implemented in `lumos-sat`.

For this work, we focus on Starlink satellites. Starlink satellites are largely comprised of two surfaces; the Solar panel and the chassis. We assume a nominal operating attitude described in [24] where the chassis points directly nadir, and the Solar panel is positioned facing the direction of the Sun shown in Fig. 3. This is known as the 'on station' configuration when the satellite has reached it's operational altitude. In this work we do not model the 'open book' configuration that is used during orbital raising, or any maneuvers that may be performed by the spacecraft to reduce brightness to observers in terminator illuminated conditions.[13] These exceptions to the nominal operating configuration will be considered in Sect. 4.1 as potential explanations for why the observed brightness of a single Starlink in our sample— Starlink-5424—varies significantly from the model prediction.

We categorise Starlink satellites by their version number and for this study target V1.0, V1.5 and V2.0 satellites. V1.5 Satellites are used in [24], and we use the surface areas of 3.65 m$^2$ for the chassis and 22.00 m$^2$ for the Solar array described in this work. Finding accurate estimates for V2.0 and V1.0 are more challenging due to the limited information SpaceX makes public about Starlink design. In a letter sent to the secretary of the federal communications commission,[14] Starlink detail surface areas for versions of the V2.0 satellites, 2 of which are to be launched from the Falcon 9 launch vehicle that are now in orbit. These are the V2.0 with Solar array surface area of 22.68 m$^2$ and chassis of 3.64 m$^2$ similarly to the V1.5 satellites, and a larger "V2.0 mini" with Solar array surface area of 104.96 m$^2$ and chassis of 11.07 m$^2$. The larger of the two hosting two Solar panel arrays.[15]

---

[13] https://api.starlink.com/public-files/BrightnessMitigationBestPracticesSatelliteOperators.pdf

[14] https://planet4589.org/astro/starsim/papers/StarGen2.pdf

[15] https://www.spacex.com/starshield/

V1.0 satellites have even less publicly available information, however rough estimates[16] place the V1.0 at 4 times smaller than the V2.0 mini, similar to the V1.5. We use these estimates to form our model in `lumos-sat`.

Due to the uncertainty around the design[17] and attitude at which the V1.0 and V2.0 satellites orbit, we exclusively use the V1.5 BRDF fitted using lab data from [24] to describe the scattering properties of the V2.0 mini and V1.0 satellites.

In order to compare our Sloan r' band results with the lumos-sat model predicted brightness, we take an ASTM E-490 Solar spectrum and integrate it with the throughput of the Sloan r' bandpass to determine a top of the atmosphere modified Solar constant of 1233 W/m$^2$. We note a caveat in this work, is that `lumos-sat` does not take into account atmospheric scattering in this model of satellite optical brightness. This work focuses on comparing observations with the existing `lumos-sat` model, and so we also do not take into account in the simulated satellite optical brightness. While out of the scope of this work, we note that this is an important direction to consider in future work to improve the `lumos-sat` model.

## 2.4 Remote Sensing Data

In order to begin exploring how the albedo and cloud properties of our observing site might impact the brightness of the satellites overhead, we use remote sensing data from the CERES (Clouds and the Earth's Radiant Energy Sysem) project[18] [38]. Single Scanner Footprint, low latency "FLASHFlux" measurements are used from the satellite Terra. With the assistance of the team at NASA's Langley Research Center, we were able to retrieve CERES Single Scanner Footprint, low latency "FLASHFlux" measurements that capture the total cloud fraction and surface albedo within the satellite's visible horizon on the surface of the Earth at the time of observation. It should be noted that CERES data is limited to conditions where the Sun is directly overhead with respect to the satellite. Data taken at a range of Solar phase angles—the angle between the Sun and the Earth as seen by the satellite—from CERES satellites is modified using pre-calculated BRDF's during the data reduction process to correct all data to conditions where the Sun is at zenith. We note again that the lumos-sat model does not include modelling of atmospheric scattering, which will impact results when comparing model residuals to remote sensing data. However, we use this study as a proof of concept to determine if future investigation is warranted, and we leave any detailed modelling of atmospheric effects to future work.

We use CERES "Shortwave" surface fluxes, which estimate total Earthshine flux over $0.3-5$ μm. In addition, surface albedo estimates are used, and cloud fractions at various atmospheric depths are combined to produce a total cloud cover estimate. We note the caveat that CERES data covers a broader wavelength

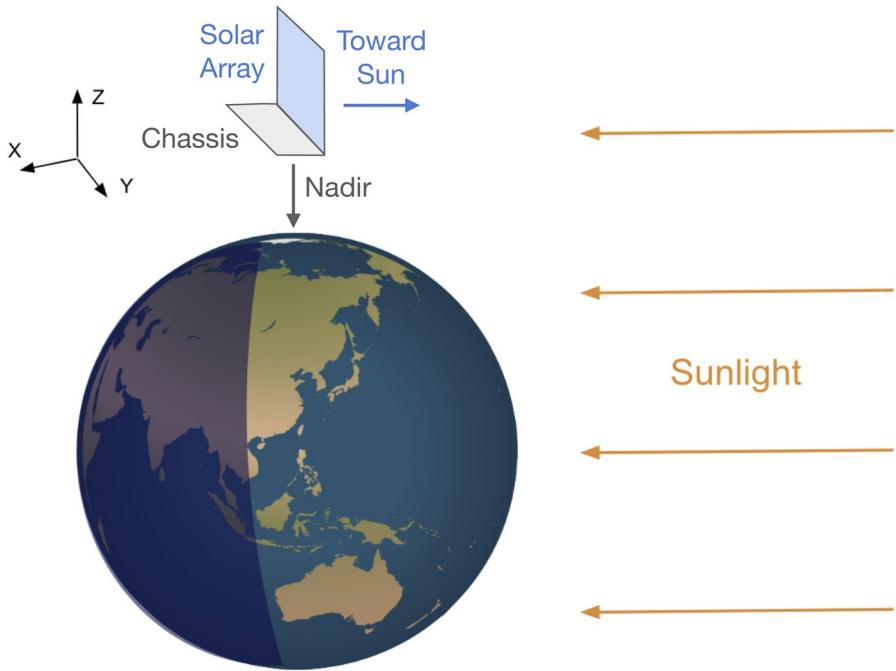

**Fig. 3** Geometry of a Starlink satellite in orbit during midday, and in terminator illuminated conditions. In nominal operations, the chassis of the satellite points nadir, and the Solar panel orientated towards the Sun

range compared to Huntsman Telescope Pathfinder r' of $\sim 0.55-0.70$ µm. We use the approach and software developed in [27] to sample CERES data at the longitude and latitude on Earth directly below the satellite of interest. Data is organised in a $1° \times 1°$ grid in latitude and longitude grid, from which values are median averaged within a calculated scattering radius $R_{\text{Earthsine}}$ based on the satellite height given by:

$$R_{Earthshine} = \cos^{-1}\left(\frac{R_{Earth}}{R_{Earth} + Alt}\right) \times R_{Earth}$$

These data are used to explore rough trends between surface albedo, cloud properties, and satellite brightness, to understand if more accurate modeling using real world data is a valid direction to explore improving satellite optical brightness models during the day. In addition, we will use this method to explore if we can determine optimal observing locations and conditions for daytime observations based on surface properties.

# 3 Results

## 3.1 Observed Satellites

We find that Starlink satellites are visible during the day using the Huntsman Telescope Pathfinder system. Of the 140 Starlink satellite passes we attempt to observe only 26% are non-detections either due to a technical fault or a non-visual detection in the region of the sky in which the satellite is targeted, leading to a success rate of 74%. This rate increased substantially as our experience and intuition improved to achieve on our last run a detection rate of 90%. We also note that we did note use Supplemental Starlink TLE data, and so there is a higher likelihood that targets that were missed may have been due to outdated TLE's. To ensure precise photometry, we rigorously filter out any data showing visual signs of high cloud cover or bushfire smoke, regardless of whether a detection has occurred. Due to back burning that took place near our observing site during one of our observing runs, this left a sample size of 81 Starlink satellites passes that can be used for photometric analysis.

Successful observations span Sun altitudes from 65° (summer, midday, at Australian latitudes) to 20°. The angle between the target and the Sun on sky spans 133° to 25°. We achieve 1447 individual observations of 28 unique V2.0 satellites, 1598 observations of 35 V1.5 satellites, and 782 observations of 18 V1.0 satellites. At our observing site at Macquarie Observatory, observations are limited in the West due to obstructions by trees to an altitude of approximately > 30°.

In Fig. 4 we present the altitude and azimuth of all of the observations made during this work that are used in photometric analysis, as well as the altitude and azimuth of the Sun at the time of observation. This plot highlights the limitations of an equatorial mount, which is the need for a meridian flip. As the telescope tracks East to West, to prevent the optical tube assembly from making contact with the pier the instrument must stop, turn 180° and recommence tracking. Due to the time that this takes, only half of every pass is captured.

We note again that the Huntsman Telescope and Pathfinder use German Equatorial mounts which have limitations in the range of satellite passes that can be tracked continuously. Starlink satellites most often travel from North-West to South-East, or from South-West to North-East. Some Starlink observations start on the West side of the pier and track until 10° past the meridian (the meridian flip line shown in green). Tracking may also commence 10–20° past the meridian flip line and travel until either obstructed by the Eastern horizon, or a visual is lost on the object. The 10–20° delay is the acquisition time for the mount to settle on the object once the tracking command is sent, and takes up a much larger angular area on sky than the Western pass due to the fact the object is traveling fastest near zenith. We should note that there is also observer bias as to which satellites and passes are chosen. Passes chosen opportunistically, not predetermined, and are chosen to reduce travel time between targets, minimise the strain on the mount (few meridian flips), and maximise the time on sky observing.

## 3.2 Satellite Daytime Brightness

We find a median observed daytime magnitude for Starlink satellites of $3.6 \pm 0.05$ mag, $\sigma = 0.6 \pm 0.05$ mag, where 0.05 mag is calculated the photometric error. This remains the same for all observations, so we do not report the photometric error for every detection. Examining the distribution of observations for each satellite version, we find that the median magnitude increases from 3.8 mag, $\sigma = 0.5$ mag for V1.0, to 3.6 mag, $\sigma = 0.6$ mag for V1.5 and 3.5 mag, $\sigma = 0.7$ mag for V2.0. The standard deviation of the distributions for each satellite version increases with later versions, indicating that they may undertake more significant variations in brightness over the duration of a pass.

Compared to twilight observations of Pomenis observatory data in [24], on average Starlink satellites are ∼ 11× brighter during the day than at twilight, and as much as ∼ 14× brighter at larger Solar phase angles. This unexpected boost in brightness means that the Huntsman Telescope Pathfinder, a modest, low-cost visible wavelength telescope has been demonstrated to detect 90% (see Sect. 3.1) of Starlink satellites that pass overhead during the day. The typical magnitude we find are higher than the median measured brightness of satellites taken at twilight conditions at Pomenis observatory of ∼ 2.6 mag, and a smaller standard deviation of 0.6mag as compared to the ∼ 1.0 mag in Pomenis data.

## 3.3 Light Curves

To determine if information about each satellite for the purposes of classification and attitude determination can be gathered from photometric light curves, we plot all observations as a function of Solar phase angle shown in Fig. 6. Larger Solar phase angles indicate a target that is close to the Sun on sky, and small Solar phase angles are targets that are far away from the Sun on sky. We also plot the twilight observations of V1.5 satellites from the work of [24] using data from the Pomenis observatory. We note that Pomenis data is not range corrected, as the intend of the work of [24] is to compare satellite brightness from the perspective of a ground based observer. Pomenis data is taken using the Johnson–Cousins V band (545 nm), which has a similar central wavelength to the Sloan r' (623 nm) used in this work (Pomenis Observatory, private communications).

Our data support the conclusion of [24] that Starlink satellites have a complicated relationship with Solar phase angle. The growing complexity of light curves with each generation of Starlink satellite may be explained by the growing complexity of the design and operation of the satellites with each new version which is further explored in Sect. 3.4. We also notice a lack of V1.5 and V1.0 satellites at Solar phase angles greater than ∼ 60°. This may be a selection bias due to our small dataset, or it could be due to a bias imposed by the daytime detection limit of the Huntsman Telescope Pathfinder as a function of sky brightness [28] due to the apparent decrease in brightness at smaller phase angles for these satellite versions. Solar phase angles greater than ∼ 60° is also where light is dominated by Solar array

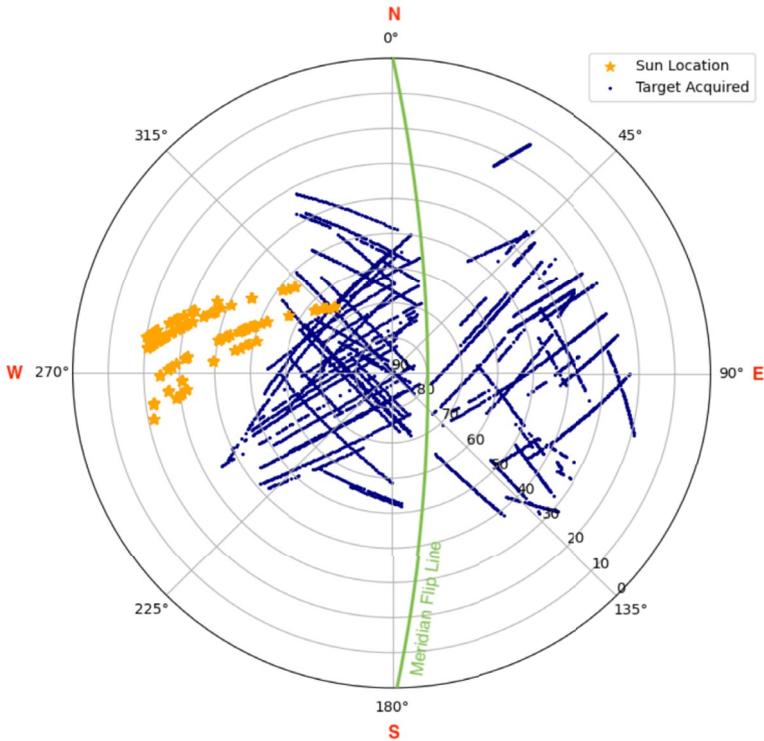

**Fig. 4** The altitude and azimuth of all successful observations of satellites used for photometric analysis. The location of the Sun during these observations is also plotted for reference in yellow. Note the approximately 30 degree minimum altitude limit in the West due to obstructions from trees. A clear division between two groups of data in the West and East can also be seen due to the limitations of an equatorial mount. We allow a liberal 10° max beyond zenith for this limitation (in the direction of travel), which is shown in green as the meridian flip line, in order to collect as much data as possible in this critical location of the pass

back-scatter [24], and so may be brighter and thus more likely to be detected against the bright daytime sky background for V2.0 satellites with two large Solar panel arrays surface area as compared to V1.0 and V1.5.

### 3.4 Differences Between Starlink Versions

We now consider individual light curves for each observed target as a function of Solar phase angle in Fig. 7. Data is organised by satellite version, and 7 data points are rolling medians over time to reduce seeing/scintillation noise due to atmospheric turbulence that is significant during the day [28]. In order to compare light curves from different passes directly, for this analysis we make a first order range correction to 500 km using a basic inverse square law scaling. We refer to these magnitudes throughout when used as "range corrected magnitudes".

In Fig. 7 the V1.0 satellites show a roughly linear trend with Solar phase angle from $\sim 80 - 140°$. A single flaring event is seen from Starlink-1906 reaching a maximum of 3.09 at 109°. Observations are on average $\sim 1$ range corrected magnitude brighter at larger Solar phase angles, perhaps due to specular chassis scatter [24]. The V1.5 satellites on average also trend linearly in brightness with Solar phase angle from 110° onwards. We examine trends in this region in more detail below. From 70° to 110° however, significant deviations from the trend is seen, with some satellites exhibiting substantial increases in brightness of $\sim 1.5$ range corrected magnitudes, while others continue a downward trend, and others reducing in brightness by $\sim 1$ range corrected magnitude. These contrasting light curves indicate the complexity of the scattering surface, may be due to varying Solar array configurations, and satellite designs. For example, modified Starlink satellites such as the "visor sat" [39] and satellites with various reflective and flocking surfaces designed to reduce brightness in terminator illumination conditions may have different reflective properties consistent with smaller Solar phase angles. We see a roughly flat trend with Solar phase angle for V2.0 satellites. Notably are two regions at 90 and 120° where large troughs or peaks are seen in the light curves, depending on the satellite observed.

While not the focus of this work, we also observe a number of non-Starlink satellites during the day, and present light curves for a selection of large bright satellites in an Appendix.

### 3.5 Comparison of Starlink Observations to Satellite Brightness Models

In order to explore why the Starlink satellites are $\sim 11$ times brighter during the day than expected, we employ the satellite optical brightness modeling package `lumos-sat` described in [24].

Using the method described in subsection 2.3, we predict the magnitude of the pass at each observation location, taking into account the different Solar panel and chassis sizes, assuming all are in a nominal operating orientation, with adjusted Solar constant to accommodate for our observation bandpass. We use the Phong BRDF parameters for the scattering function of the surface of the Earth given by [24] for the ocean, due to the fact that they spend a majority of the duration of the pass over the east coast of Australia. These are $K_d = 0.48$ for diffuse surface albedo, $K_s = 0.08$ for the magnitude of the spectacular component, and n = 16.45 for the width of the specular peak. We assume the same BRDF profile generated using in lab testing in [24] for the V1.5 satellites for all versions in this work, as we do not have access to any further information about the scattering properties of V1.0 or V2.0 satellites.

The model is run twice to calculate magnitudes with and without including an Earthshine component, and the residuals are shown as histograms in Fig. 8 for each satellite version. We remind the reader that only V1.5 satellites are using an informed spacecraft design, while for the other versions we had to make assumptions about their physical properties (see Sect. 2).

Considering the two models in Fig. 8, the model that does not include Earth-shine has a bimodal distribution. This is indicating that some passes are dominated by Sun only illumination (which are well described by the model), while others are Earthshine dominated. This is explored further in Fig. 9 where we show the residuals for each model plotted as a function of solar phase angle. We can clearly see from this plot that the Sun only model cannot explain the sources of illumination of Starlink satellites during the day for solar phase angles greater than $\sim 80\,°$. We note a small number of V1.5 data points that are not well described by either model from Starlink-5424 which is investigated in further detail in subsection 4.1. We also note a limitation of the Sun only `lumos-sat` model, that for some passes with solar phase angles $\sim > 100$ the model predicts no sources of illumination, and so sets an arbitrarily small magnitude at 12.5 for these points, indicated by the downwards facing arrows in Fig. 9. The model including Earthshine describes that data well, and we find residuals of 0.89 mag, $\sigma = 0.82$ mag for V2.0, 0.4 mag, $\sigma = 1.01$ mag for V1.5 and 0.73 mag, $\sigma = 0.37$ mag for V1.0.

The difference between model predictions shown in Figures 8 and 9 indicates that light scattered off the Earth's surface is a significant factor in the brightness of some satellite passes during the day, in particular for solar phase angles $\sim > 80\,°$. This is similar to the finding of [24] for observations of satellites in terminator illuminated conditions after Sunset. We find that the model best describes the brightness of V1.5 satellites, which is consistent with using the V1.5 surface BRDF (which was provided by SpaceX) for these calculations.[19]

For the other satellite versions, we note again that we had to make assumptions about their physical properties. The V2.0 satellite model has the largest variation from the model, and is fainter than calculated in the model. This may be due to an overestimation of the size of the Solar panel array, or improved brightness reduction surfaces. Not all V2.0 satellites may have the same solar panel array, and so may less scattering surface area than is used in this model [40]. However, SpaceX does not provide information about which satellites are in what configuration. Additionally, the scattering surface of the satellites could be less reflective than the V1.5 satellites.

To further explore possible physical differences between each version, we isolate a set of "control sample" data and present in Fig. 10 light curves and transit paths of two satellites of each version. The control sample satellites have the largest number of individual observations and largest Solar phase angle range. Black data points have a mean rolling window of 7 observations, and light grey points are individual observations. The red dotted line is the predicted magnitude of the target from the model that includes Earthshine. For this comparison of light curves, we again make a first order range correction to 500 km.

For the two V2.0 satellites we see similarities in the light curve with a peak at 114°, followed by troughs at different Solar phase angles. The two satellites are transiting in front of the Sun in mirrored paths, but similar Sun altitudes. This may explain the variation in light curves as different sides of the chassis is visible to the observer at this phase angle. V1.5 and V1.0 do not show these statistically significant peaks and troughs. However we do note qualitatively that V1.5 satellites do appear to be more complex than V1.0.

---

[19] Excluding the single outlier satellite Starlink-5424 investigated in further detail in the discussion.

Model light curves show similar rise or small of the data, but do not capture the complexity of the light curves. The lack of complexity in the predictions is likely due to the simplicity of the scattering surfaces modeled, which does not contain any information about self shadowing of components, differences in scattering properties on different parts of the Solar array (front vs back for example) or the chassis. We find again that V2.0 models overestimate the brightness of the target, V1.5 are a good order of magnitude fit, and V1.0 models are underestimated. Again, this reflects the need to make assumptions about V1.0 and V2.0, whereas V1.5 models were provided by SpaceX in [24].

To broadly compare the models to our observations and investigate the impact of Earthshine, we compute the all sky satellite magnitude in transit path plot using models with and without Earthshine, for the V1.5 satellites in Fig. 11. As described in [24], the plot can be thought of as the sky above an observer on the ground looking up, with zenith at 90° in the center of the plot, stepping down in units of 10° to the horizon along the circumference of the plot. It can be used to predict where it is most likely to see a satellite during the day.

Figure 11 shows that the model without Earthshine does a poor job explaining the data, particularly for large Solar phase angles. The bright spot in these plots corresponds to Sunlight specular reflection off the Solar array at small Solar phase angles, which decreases in brightness as the Sun sets. The model cannot explain the brightest observations at larger Solar phase angles, where the satellite is directly overhead from the observer.

The model that incorporates Earthshine better matches the observations. The data seem to support the idea presented in Fankhauser et al. [24] that there is a bright Solar panel spot opposite the Sun at large phase angles due to Earthshine, as well as a diffuse and specular Earthshine scattering component off the Chassis.

### 3.6 Investigation of Cloud Fractions and Surface Albedo Variation on Satellite Brightness

We have found scattered light from the surface off the Earth has a considerable impact on the brightness of satellites observed during the day. We now explore some of the factors that influence the intensity of Earthshine, and if there are conditions under which satellites appear brightest. We explore the impact of surface albedo on the brightness of the satellites observed during the day by correlating satellite observations with Earthshine flux calculated using CERES satellite weather data taken on the same day and location to when the satellite is observed overhead from the ground, however covering a broader wavelength range (see Sect. 2.4). This information may be useful in designing future iterations of daytime optical satellite brightness models.

This analysis also aims to explore the idea presented in [6] that clouds surrounding an observing site may boost local albedo and thus satellite optical brightness, as long as the clouds are not directly above the observing sight. This is possible due to the relatively large scattering area beneath a LEO satellite. For Starlink satellites orbiting at a mean altitude of $\sim 560$ km, this is a circular region projected onto the surface of the Earth ($2 \times R_{\text{Earthshine}}$) of 5463 km—larger than the diameter of the continent of Australia. Cloudy local areas that have observing sites above the clouds (such as mountain top observatories) may be optimal for daytime SDA, and investigations may be useful in identifying optimal locations at which optical daytime SDA facilities should be built.

Using the method described in Sect. 2.4, we compute the median surface albedo beneath a satellite for observations in two groups: observations where satellites are within the area of the sky which is dominated by specular chassis scatter (directly overhead), and an area dominated by Solar array backscatter (opposite the Sun). Maps of CERES albedo from the top of the atmosphere which encompasses both changes in surface albedo from different types of landmasses and oceans as well as cloud formations are shown in Fig. 12. We note that this measure of albedo is restricted to midday observations by CERES satellites where the Sun is directly overhead. We do not attempt to scale the albedo by assuming a BRDF (scattering function) for the surface of the Earth to accommodate different geometries.

In Fig. 13 the model residuals (measured - predicted satellite brightness) is examined as a function of the CERES albedo beneath the satellite at the time of observation. A linear fit is made to observations of V1.5 Starlinks for which we have the best estimates of their physical properties to help indicate the presence of a trend. For observations in a region of the sky that are Sunlight dominated (small solar phase angles $< 100°$ from Fig. 9) there is no clear linear relationship with albedo. Residuals follow the same features as shown in Fig. 8, where Starlink V2.0 satellites are fainter than predicted. This may indicate that it is the scattering properties and dimensions of the chassis of the Starlink's that are the dominant contributing factor to the model discrepancies between satellite versions with observations. This may also reflect limitations of CERES data, due to the fact it covers a broader wavelength range and is not immediately coincident with satellite observations. In addition, CERES albedo is calculated as an average over the entire scattering areas and does not taking into account the orientation of the satellite with respect to the observer.

When considering observations in regions of the sky where Earthshine is the dominant source of illumination (large solar phase angles $> 100°$), we do see a relationship with surface albedo in Fig. 13. The model generally overestimates the brightness of satellites that are located over regions with lower median albedo, however there is a large amount of scatter. The relationship may be more pronounced than in observations that are chassis dominated, due to the much larger reflective surface area of the Solar panel arrays ($\sim 6\times$ larger for V1.5), and differing scattering properties compared to the chassis.

To further explore this, we take observations of V2.0 satellites that have the largest Solar panel surface area, and compute the median brightness for observations with an albedo in the lower 1% percentile with median albedo = 0.41 and the upper 99% percentile with median = 0.49. We find that the median observed brightness of satellites above areas with $\sim 8\%$ higher albedo are $\sim 1.6\times$ brighter at $3.36 \pm 0.0004$ mag (where 0.0004 is the standard error on the median) than for observations over areas on the Earth of lower albedo at $3.87 \pm 0.0002$ mag.

As a caveat on the above analysis, we note the albedo ranges used are relatively limited. In addition to gathering data with a larger albedo range, in future it may be worth weighting regions on the Earth's surface based on the direction the satellite is facing with respect to the observer, and how close the clouds are located to the observing horizon, in comparison to the conditions directly below the satellites.

The results shown in Fig. 13 may indicate that it may be worth considering the albedo of the Earth's surface and density of cloud cover around an observing locations that are dedicated to daytime optical space domain awareness efforts, however more data is needed in a wider variety of observing conditions to confirm this tentative result.

## 4 Discussion

In this work we have shown that Starlink satellites are clearly visible during the day with the Huntsman Telescope Pathfinder, and are on average $\sim 11\times$ brighter than similar observations taken at twilight. We find that the median brightness of Starlink satellites during the day is 3.6mag, $\sigma = 0.6$ magnitudes. Our results are of a similar order of magnitude to the findings of [9] who report detections of Starlink satellites between 3.5–4 magnitudes and some up to a brightness of 2.6 magnitudes in a 570nm long pass filter. Our findings reflect this magnitude range, and in particular satellites with brightness exceeding 2.6 magnitudes in Sloan r' for short periods of time.

We show that this marked increase in brightness is likely due to the contribution of diffuse Earthshine scatter beneath the satellite at the time of observation by comparing our results to satellite optical brightness models. The contribution of Earthshine that boosts the brightness of satellites may make them accessible to be observed during the day at optical wavelengths by modest SDA facilities build from off the shelf components, like the Huntsman Telescope Pathfinder, and eventually the Huntsman Telescope [34].

We have also demonstrated that the Huntsman Pathfinder is capable of producing light curves of Starlink satellites. We have shown that these light curves are complex and vary between satellite versions. We now discuss how these light curves taken during the day may be useful when monitoring satellites for SDA applications.

### 4.1 Investigation of Observed Outliers

There are multiple factors that impact the brightness of satellites, and contribute to their complex light curves. These may include the orientation of the satellite during the pass, it's physical size, structure and the materials that it is made of, as well as the non-uniform scattering of light off the Earth's surface (e.g., [11, 14, 24]). A possible explanation for the variations seen in Figs. 7 and Fig. 10 may be due to complex surface features on the body of the satellite changing the way that light is scattered from the surface, as well as the possibility of self shadowing raised by [24] and discussed in [14].

Another explanation is that the satellite might be flying in an unexpected orientation. By comparing model predictions to daytime satellite light curves, we explore if it is possible to determine if outlier observations could be explained by this hypothesis. During our survey, a single V1.5 satellite, Starlink-5424, was observed on the 9th March 2024 with a magnitude range of 1.99 mag to 4.26 mag over a solar phase angle of 78 to 88°. Interestingly, the magnitude observed is 3.55 magnitudes brighter than the model predicts if the satellite where to be orbiting in a nominal configuration. This is unexpected, as we have good estimates for the surface area and scattering properties of V1.5 satellites, and all other V1.5 satellites are shown in Fig. 5 to match the model within ∼ 0.4 mag. In addition, a visual inspection of the observations confirms the presence of a satellite in the data. The satellite was launched in December of 2022, and there is no indication that it is special in anyway, or if it is de-orbiting, and its status (at the time of writing, June 2024) is operational.[20]

If we consider that the object is indeed a Starlink satellite in the standard V1.5 configuration, the large discrepancy in predicted brightness may be explained by a change in the orientation of the satellite that is not nominal (i.e., solar panel always facing the Sun—see Fig. 3). To investigate this, by trial and error we rotated the model satellite's chassis normal vector and Solar array normal vector to minimise the residuals at the location of the sky matching our observations. We find one configuration that matches the brightness offset. The best configuration is when the Solar panel would have to be rotated 40° away from the Sun. The resulting polar sky chart and satellite transit is shown in Fig. 14. We note that Solar panel specular reflection is rotated from the nominal position shown in Fig. 11 as a result of the change in Solar panel orientation.

Why would a Starlink adopt such an orientation? This may be an example of an orbital maneuver attempted by SpaceX, whereby the Solar panel orientation is altered when crossing into terminator illuminated conditions[21][22][23] as Starlink-5424 was doing at the time of observation. This is an effort to reduce the brightness of the satellite to observers on the ground when the satellite is visible at twilight, at the expense of optimal Solar array orientation for power generation. This example hypothesizes the potential for the model, coupled with optical daytime observations to determine the orientation of a satellite in orbit, without resolving it. However it is important to note that many more observations of this object over a large range of Solar phase angles is needed to confirm this hypothesis.

---

[20] https://satellitemap.space/?norad=52864

[21] https://spacenews.com/spacex-to-test-starlink-Sun-visor-to-reduce-brightness/

[22] https://www.tesmanian.com/blogs/tesmanian-blog/spacex-visorsat

[23] https://skyandtelescope.org/astronomy-news/the-newest-and-largest-starlink-satellites-are-also-the-faintest/

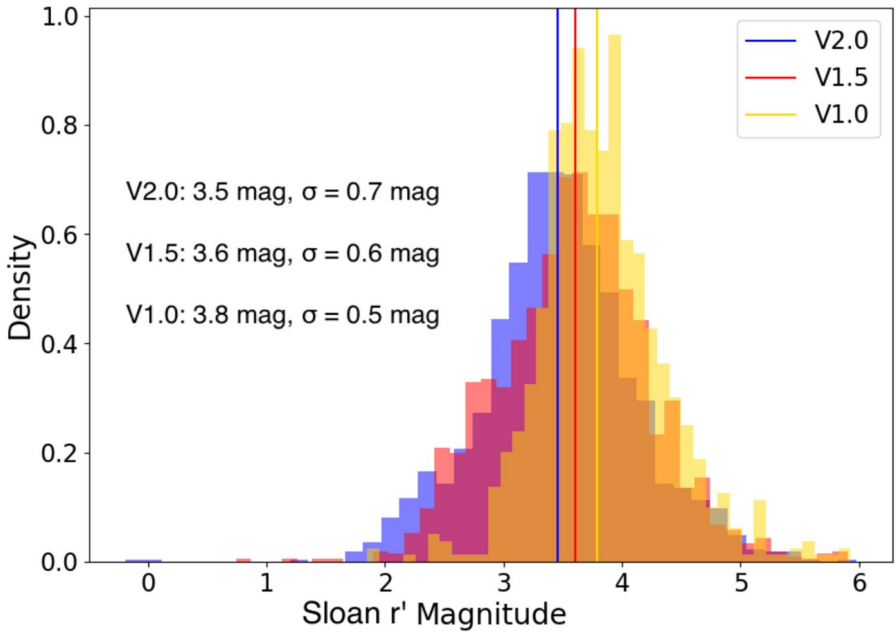

**Fig. 5** The Sloan r' magnitude of Starlink satellites observed during this survey. We include individual measurements from every exposure, and organise the data by satellite version. Considering all satellites, the median daytime r' magnitude is 3.6mag, $\sigma = 0.6$ mag. For each satellite version, we find that the median magnitude increases from 3.8mag, $\sigma = 0.5$ mag for V1.0, to 3.6mag, $\sigma = 0.6$ mag for V1.5 and 3.5 mag, $\sigma = 0.7$ mag for V2.0. We also find that the standard deviation of this distribution increases for later models as well, indicating that the V2.0 satellites are on average brighter, and more significant variations in brightness over the duration of a pass. We note that there may be selection bias on the right hand side of the distribution due to our daytime detection limit, which is the most restrictive at V band 4.6 AB mag [28]

## 5 Future Work and Improvements

The success of this pilot survey opens opportunities for improving observational techniques, data quality and quantity for future observations. Here we briefly share future work and improvements that will be implemented to our existing pathfinder instrument, as well as an autonomous mode on the Huntsman Telescope.

Firstly, restrictions imposed by the equatorial mount used in this study, as well as obstructions on the horizon, prevents us from full light curves of targets. The importance of observing a large range of Solar phase angles for a single target has been discussed in relation to object classification and assessment of satellite orientation, and so the move to a fork mount or similar is a natural step towards creating a dedicated daytime SSA facility.

There are also several methods that can be employed to reduce the impact of seeing/scintillation which is explored in detail in [28], and also impacts the quality of satellite observations during the day. One method is to increase data rates, such that many observations can be averaged without significant expensive of temporal

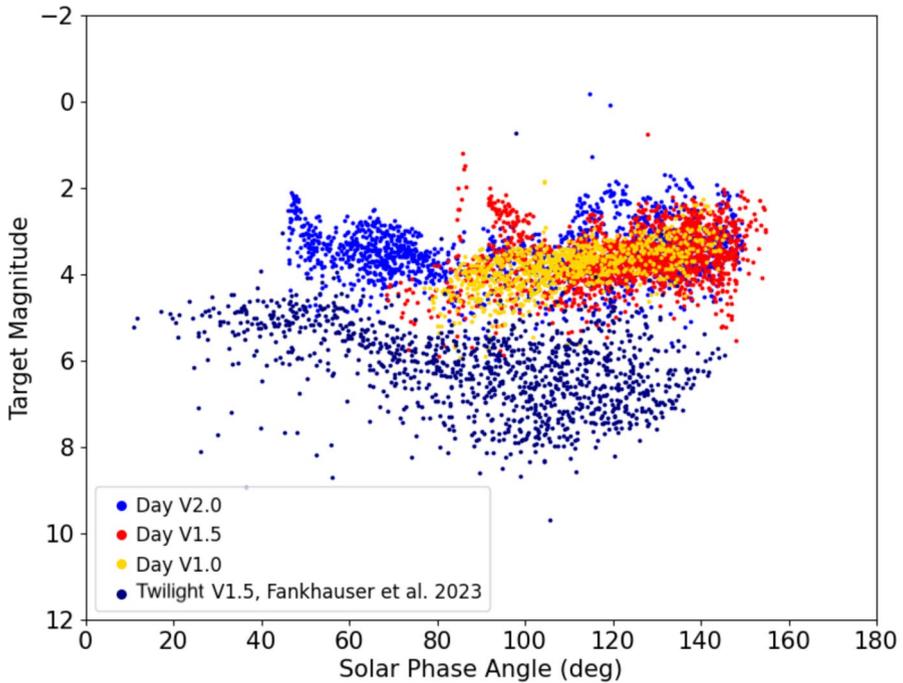

**Fig. 6** Daytime observed target magnitudes for Starlink satellites in Sloan r' as a function of Solar phase angle. V1.0 are shown in yellow, V1.5 in red and V2.0 in blue. Pomenis observatory twilight observations of V1.5 satellites from [24] are shown in navy. We note that the relationship between Solar phase angle and target magnitude for Starlink satellites is complex, and varies between satellite versions. The median brightness is also 2.6 magnitudes brighter during the day than twilight observations. We also note a lack of observations of V1.5 and V1.0 satellites at Solar phase angles greater than ∼ 60 ° which may be due to observational or detection limit biases

resolution. Onboard data processing that makes use of the Huntsman Telescope's Nvidia Jetson computers, and optimising data transfer speeds, is a potential next step to increasing data rates.

The Huntsman Telescope also consists of multiple lenses which enables it to image in multiple bandpasses simultaneously. This capability opens up the opportunity of obtaining satellite colour information, which may also help in object identification and monitoring. Finally, automating observations by incorporating the daytime observing workflow into the Huntsman Telescope ecosystem, including autonomous collection of bright reference stars during the day and intelligent sensor tasking, will improve photometric calibration due to the larger sample size, and maximum number of targets observed throughout the day.

Finally, updating the Lumos-sat model to include atmospheric radiative transfer modeling, particularly for Earthshine dominated observations, is needed to improve estimates of the magnitude of satellites during the day. To better understand the impact that Rayleigh scattering and atmospheric absorption has on the accuracy of the existing model, we make an order of magnitude estimate by taking a standard reference solar spectrum ASTM G-173 at the bottom of the atmosphere and inte-
grate it with the Sloan r' bandpass used in this work to derive a modified solar con-
stant. This is used to compare with model estimates made using the ASTM E-490
Solar spectrum that is used in Lumos-sat at the top of the atmosphere. Using this
modified solar constant, we recompute the brightness of satellites with a solar phase
angle > 110°, or those observations most likely to be dominated by Earthshine as
shown in Fig. 9. We find a median decrease in magnitude of 0.32 fainter than the
Lumos-sat model predicts, confirming that this limitation of Lumos-sat needs to be
address for future, more accurate estimates of satellite brightness. Further we note
that in future, more detailed modeling, it is likely that this effect is larger due to mul-
tiple passes of the scattered light through the atmosphere.

## 5.1 Can Satellite Attitude be Determined from Light Curves?

As part of this future work section, we make some order of magnitude calculations
to explore the possibility for recovering satellite attitude from ground based, daytime
light curves. In particular, we look for variation that is orders of magnitude fainter
or brighter than the model prediction due to a satellite orbiting in an unexpected
orientation (i.e. it has lost attitude control) and determine if these variations can be
detected by the Huntsman Telescope Pathfinder during the day. Again, we stress that
this is only an order of magnitude calculation used to guide future direction for this
work, and not a detailed study of the ability to infer attitude from light curves alone.
If the magnitude of variation in satellite brightness at different attitudes is << less
than the photometric accuracy of the Huntsman Telescope during the day, then this
is not a direction for future work. Inferring satellite attitude from light curves alone
is a complex task (see e.g. [14]) and is well beyond the scope of this work.

In Fig. 15, we explore the impact of a change in satellite orientation with respect
to the Sun, observer, and the Earth. We use an example satellite control sample, a
V1.5 target, Starlink-5915, whose light curve is shown in Fig. 10 and shows a simi-
lar observed magnitude as the model predictions. The flight path, and orbital altitude
of the satellite is kept identical to that observed in Fig. 10. We explore two scenar-
ios, one in which the solar panel is rotated about the Z axis, such that it faces away
from the Sun at various ° away from nominal. We also explore a tumbling scenario,
where the satellite is rotated about the X axis which is identified as being of interest
for SpaceX SDA capabilities [41].

In the rotation scenario, the model predicts subtle variation in the light curve
brightness for this particular pass that is most apparent at small Solar phase angles
below 100°, in the Solar array back scatter region. Variation in brightness are found to be 0.74mag, $\sigma = 0.43$ mag. This change is due the variation in Solar panel surface area and therefore mostly scales by the amount of diffuse reflection as seen by the observer as the panel turns. We also see a peak at ∼ 140 °, where the satellite is closest to the Sun in its pass. This is due to a specular reflection off the Solar panel with respect to the observer and the Sun at these particular configurations.

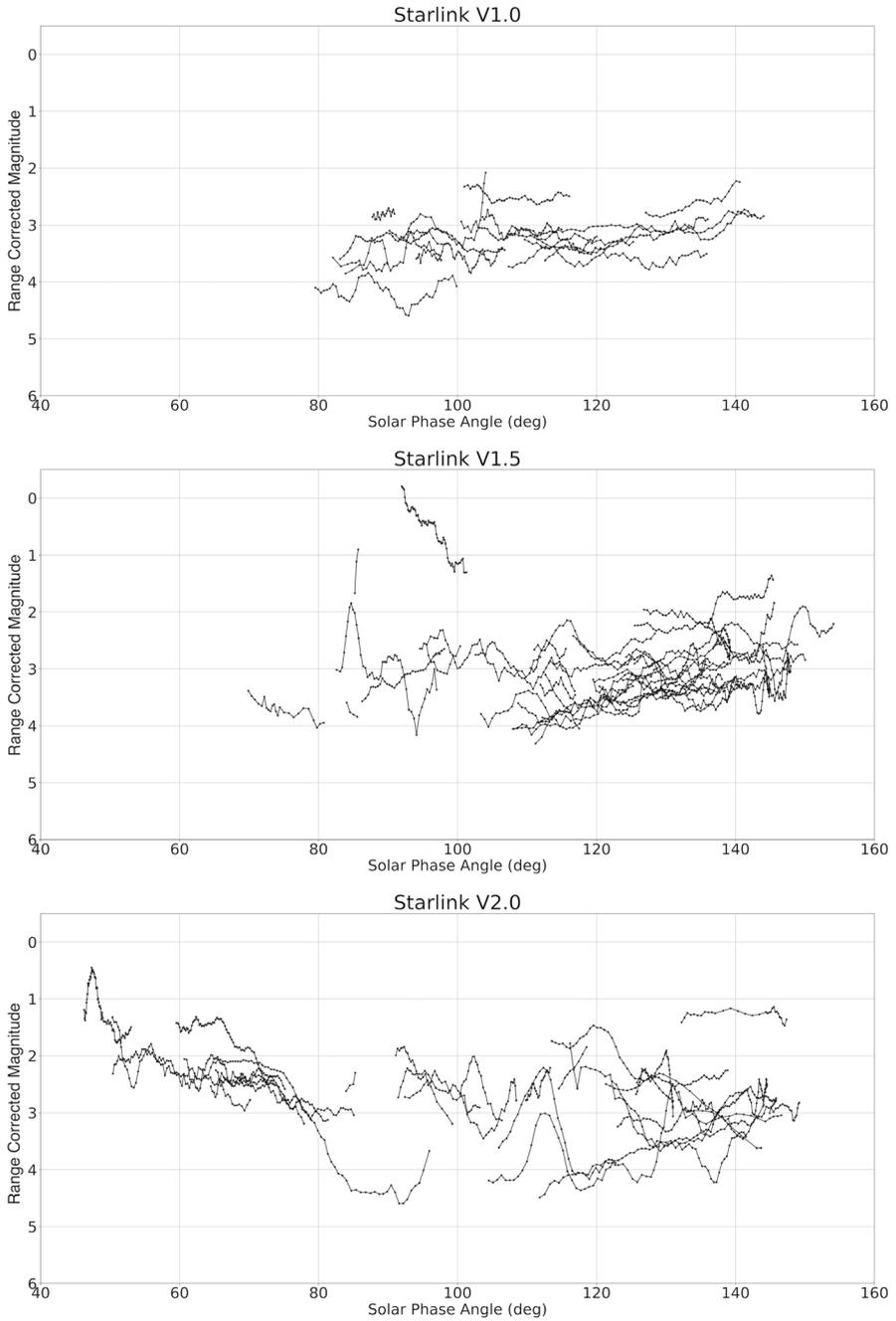

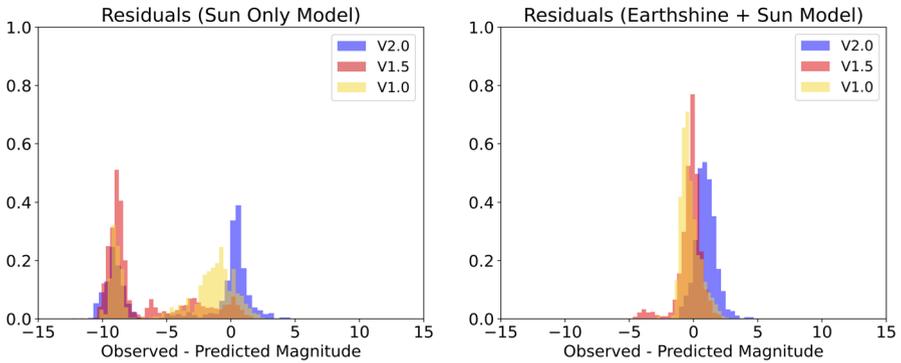

**Fig. 8** Predicted Sloan r' target magnitude subtracted from observed target magnitude for two model configurations. The left panel shows residuals for a model without including a light source due to Earthshine, and the right includes Earthshine. We find that the model that best fits the data is that which includes contributions due to light scattering off the surface of the Earth, in agreement with a similar conclusion for observations during terminator conditions [24]. Our work indicates the Earthshine is a significant contributors to the brightness of satellites during the day. For the model including Earthshine we find mean residuals of $0.89, \sigma = 0.82$ mag for V2.0, $0.4, ;\sigma = 1.01$ mag for V1.5 and $0.73, \sigma = 0.37$ mag for V1.0. Discrepancies between observations and the model including Earthshine are likely due to a differing scattering properties (BRDF) for the satellite or the size of satellite surface areas. In particular we note that V2.0 observations are generally fainter than model predictions, indicating that the surface area assumed in the model may be over estimated. We also note a small number of V1.5 observations with negative residuals, which are attributed to Starlink-5424 and is explored in more detail in Sect. 4.1. The model without Earthshine produces a bimodal distribution, with observations that are dominated by direct solar illumination having smaller residuals than observations that are dominated by Earthshine. This is explored further in Fig. 9

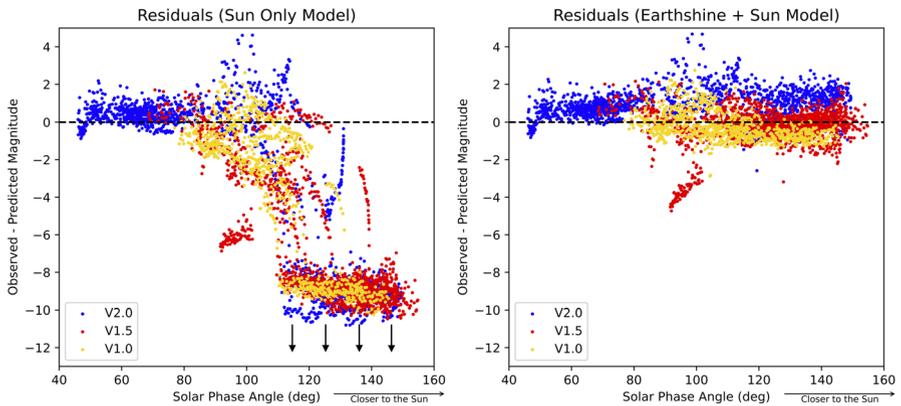

**Fig. 9** Predicted Sloan r' target magnitude subtracted from observed target magnitude plotted as a function of solar phase angle for a model only including illumination from the Sun, and a model including Earthshine and the Sun. For the Earthshine + Sun model we see generally good agreement for each of the Starlink models. V1.5 residuals that do not follow the trend at a solar phase angle of $\sim 100$ are from Starlink-5424 and will be further investigated as an outlier. For the model excluding Earthshine, we see good agreement for small solar phase angles where the Sun illumination is the largest contributing factor from the point of view of a ground based observer. For larger solar phase angles where an Earthshine component is expected to dominate, the model does not describe the data well. So some data in the Sun only model, `lumos-sat` predicts no flux, and so sets an arbitrary small value of 12.5 magnitudes. This is indicated by the downwards point arrows. It is clear from this figure that the model including Earthshine better describes the sources of illumination of Starlink satellites during the day

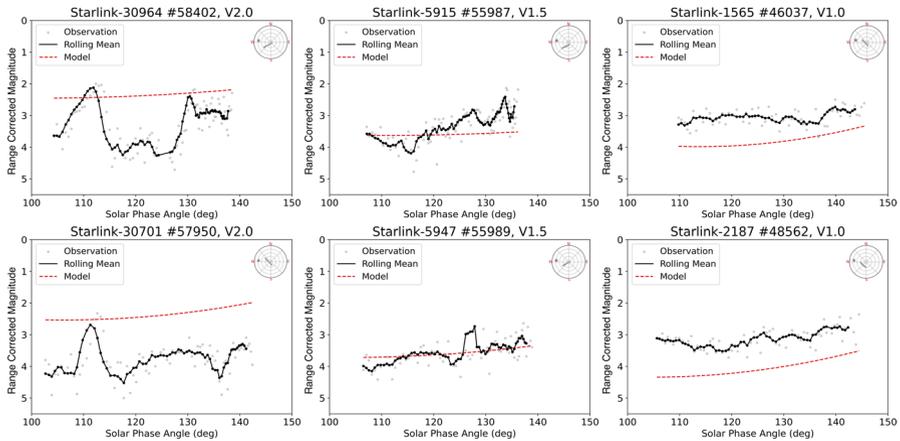

**Fig. 10** "Control sample" satellite light curves for 2 of each satellite version with highest number of consecutive observations. Starlink satellite number, NORAD ID (#), and Starlink version are given in the heading. The magnitude of the target in r' band is presented as a function of Solar phase angle, in addition to the transit path in the top right hand side of each plot. Black points are rolling means (7 observations) with time to reduce the effect of seeing, with all observations plotting in light grey. The red dotted line is the predicted magnitude of the target from the model. The location of the Sun is plotted as a star, and satellite locations as grey circles in the top right transit path plot. The light curves, and modeled satellite brightness presented in this figure have both been range corrected to 500 km in order to compare passes. We notice some similarities, notably a peak at ∼ 115 ° phase angle, and a double trough features for the V2.0 satellites at higher Solar phase angles. The model does not capture the complexity of the light curves, but instead offers a rough order of magnitude estimate for target brightness. V2.0's are over-estimates in brightness, V1.0's are under estimated, and V1.5's are the best fit, as we have information about the scattering properties and surface area from [24]

For the tumbling scenario, the model predicts a variation in brightness of 2.7mag, $\sigma = 2.04$ mag as the entire satellite is rotated. Rotations between 135° and 225° resulting in the satellite being fainter than the approximate Huntsman Pathfinder magnitude limit of ∼ 6 mag (in the evening) [28] and so is not included in this plot. From these simulations we can conclude that in most scenarios in this order of magnitude calculation, the variation in brightness from the nominal light curve is within the photometric accuracy of the Huntsman Pathfinder during the day of 0.05 ± 0.03 mag [28]. These simulations indicate that exploring the ability to recover satellite attitude from photometric light curves during the day may be a potential avenue for further investigation.

## 6 Conclusions

In conclusion, we find that Starlink satellites are visible during the day using the Huntsman Pathfinder system. By comparing results to satellite optical brightness models, we find that the surprising increase in satellite brightness of up to ∼ 11×

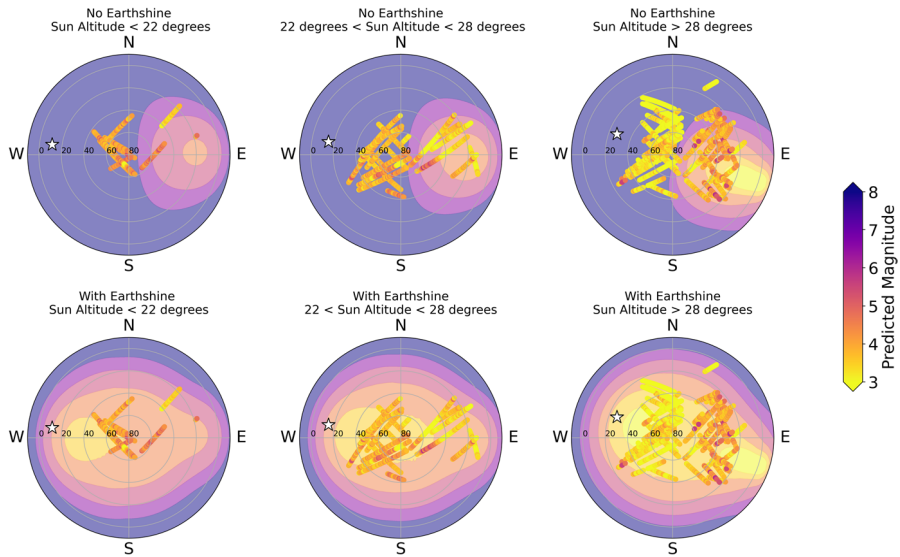

**Fig. 11** Predicted V1.5 satellite magnitude across the whole sky for low, medium and high Sun altitudes, with observed V1.5 satellite brightness over plotted for all targets. The colour bar is the target magnitude. The median Sun location is shown with a white star. Zenith, or 90° altitude to the perspective of an observer on the ground is the center of the plot, stepping down to the horizon in units of 10° to the circumference of the plot. The top 3 panels show the predicted magnitude without Earthshine included as a light source, and the bottom 3 panels include Earthshine in the model predictions. We find that the model including Earthshine better matches the observed distribution of satellite brightness at large Solar phase angles. Starlink-5424 is shown on the far right plots as an outlier (top right quadrant), and is addressed separately below

during the day as compared to twilight observations is likely due to the contribution of Earthshine below the satellite during the day.

Starlink satellites are found to have a median brightness of 3.6mag, $\sigma = 0.6$ mag in Sloan r', with later Starlink versions found to be brighter on average than earlier ones, likely due to their larger Solar panel and chassis surface area. We also present daytime optical light curves for 81 Starlink satellites, and conclude that they are complex and are generally similar across Starlink versions.

We find an order of magnitude agreement in observed satellite optical brightness models that include an Earthshine component of 0.89mag, $\sigma = 0.82$ mag for V2.0, 0.4mag, $\sigma = 1.01$ mag for V1.5 and 0.73mag, $\sigma = 0.37$ mag for V1.0. The largest deviations from the model are found for versions which we do not have information about the scattering properties of the chassis and Solar panel, or accurate information about the surface area. V1.5 satellites for which we do have this information, we find generally good agreement with the model. In comparison for models that do not include Earthshine during the day, we find a bimodal distribution, and significant deviations from the modeled brightness for satellites at solar phase angles greater than $\sim 80°$. From these results we conclude that the brightness of satellites observed

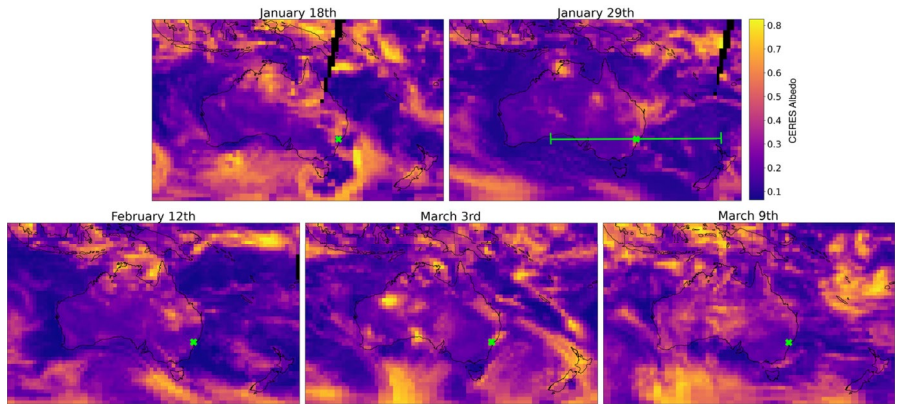

**Fig. 12** Albedo from the top of the atmosphere as seen by CERES satellites in 1 × 1 degree grids for the days on which Starlink satellites are observed from the ground. The observing location at Macquarie University Observatory, Sydney, Australia is shown as a green cross. For illustration the approximate diameter of the scattering area beneath a Starlink satellite at altitude ∼ 560 km is shown in the top right panel in green. Pixels that do not contain data for that particular day are shown in black. Averages are taken for albedo beneath the satellite and cover large swathes of the Earth, and so are unlikely to impact the results. Observing conditions span a large range from largely clear days such as February 12th and March 9th, to observing in between cloud bands such as January 18th. Albedo measurements vary between ∼ 0.25 − 0.35 for arid areas and vegetation, to ∼ 0.10 over cloud free ocean, and ∼ 0.80 in thick cloud bands

at large solar phase angles during the day can only be accurately described using a model that includes both an Earthshine and Sunlight component.

We also find that the light curves are more complex than those computed by simple models that only take into account two uniform scattering surfaces. It is likely that variation in scattering properties on different components of the chassis, as well as further complexities like self shadowing, contribute to more complex light curves that are observed with the Huntsman Telescope Pathfinder.

We find that there is a potential relationship between residuals of model satellite brightness and observations as a function of albedo on the surface of the Earth, which may indicate the need for incorporating real time weather data into future daytime satellite brightness models in order to improve their accuracy. It may also be useful to consider when determining the best sites to position optical daytime SDA facilities.

As part of our investigation into future work, we explore an order of magnitude calculation to determine if a change in satellite orientation can be detected during the day from satellite light curves. We determine that for some scenarios—namely out of control tumbling of the satellite—the variation in brightness is larger than the photometric accuracy of the Huntsman Telescope during the day. As a result, we conclude that this may be a possible avenue for further investigation in the future.

These findings demonstrate the potential for small optical telescopes built predominately from off the self hardware to make a meaningful contributing to Space Domain Awareness efforts. This is particularly useful in the upcoming era of mega constellation such as AST's Space Mobile (BlueWalker 3), Eutelsat's OneWeb,

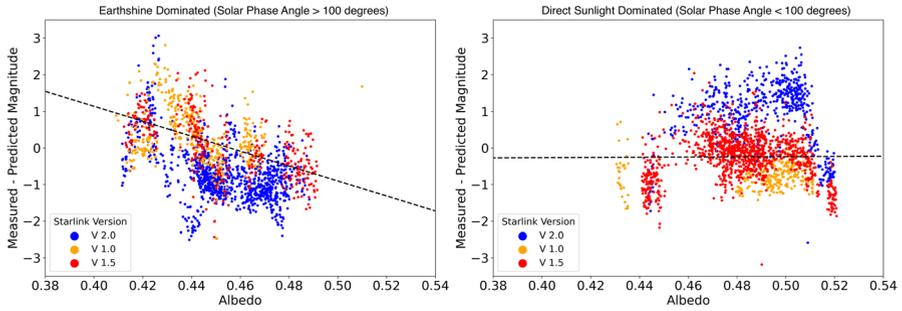

**Fig. 13** Model residuals (measured − predicted satellite brightness) for each satellite version as a function of median Earth albedo seen by a satellite as estimated using CERES observations on the same day that satellite observations were carried out. A linear fit is shown for V1.5 satellites for which the physical properties of the satellites are well known to help indicate a trend visually. The data is split between observations that are Earthshine dominated (large solar phase angles > 100) and Sunlight dominated (smaller solar phase angles < 100). We find that only observations that are Earthshine dominated show a relationship with albedo, where observations over areas with lower surface albedo are generally fainter than the model predicts

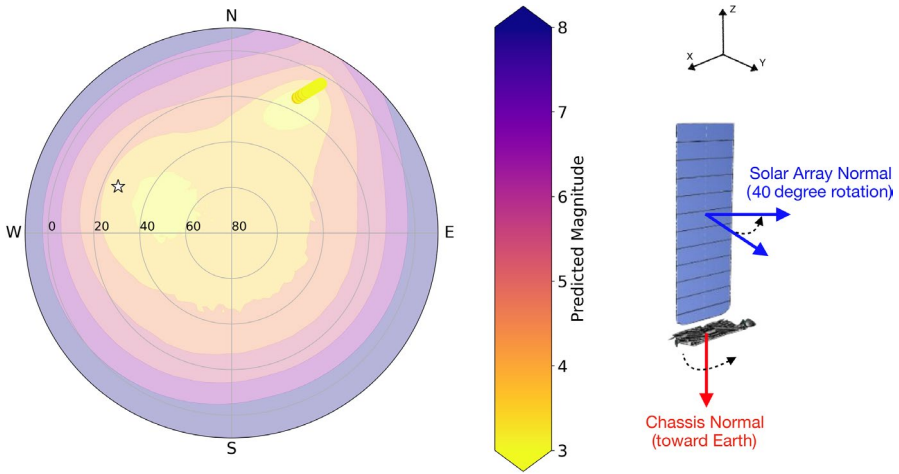

**Fig. 14** Polar sky plot predicting the magnitude of a satellite as a function of position on sky for a V1.5 design at the time at which Starlink-5424 is observed. The Solar panel array, which is nominally assumed to be facing the Sun, has been rotated in the model by 40° away from the Sun in order to match the brightness of the satellite observed. This scenario may be consistent with an attitude adjustment described by SpaceX in order to reduce the brightness of the satellite in terminator illuminated conditions

Amazon's Project Kuiper, and Starlink. With the exponential growth of Starlink satellites in orbit in recent years and the planned launch of many more - particularly as we are enter the Starship generation—the ability to monitor and produce photometric light curves of starlink satellites during the day is highly desirable. This capability has the potential to make the Huntsman Telescope a highly productive SDA facility. Daytime SDA capabilities of LEO satellites will reduce the dependence on the short observation windows during terminator illuminated conditions, allowing for improvements in satellite tracking and attitude monitoring, situational awareness and object classification.

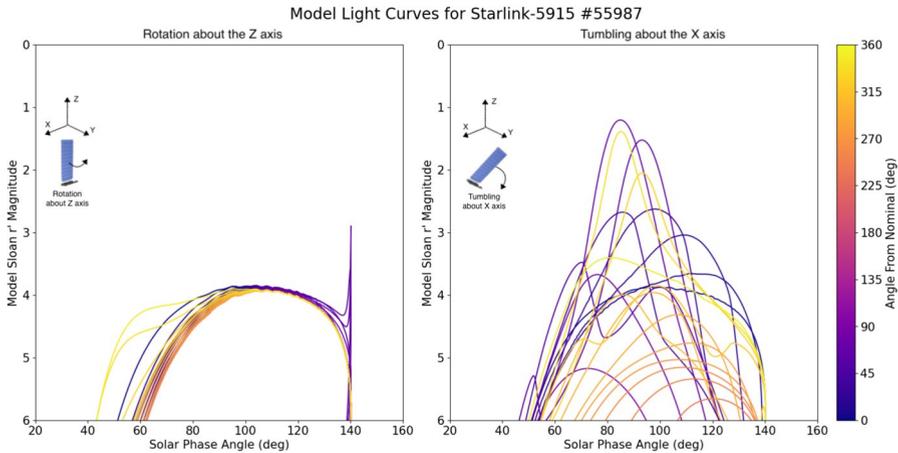

**Fig. 15** We explore the impact of rotating a satellite on the predicted light curve as a function of solar phase angle for various deviations from nominal. The panel on the right shows the results of a tumbling simulation about the X axis for Starlink-5915, following the same path as shown in Fig. 10. The Y axis is cut at 6 magnitudes which is a reasonable detection limit for the Huntsman Pathfinder in optimal daytime observing conditions. Passes between 135 and 225° result in light curves fainter than 6 magnitudes and so are not shown. Large deviations in brightness from the nominal configuration are seen with a mean of 2.7mag, $\sigma = 2.04$ mag, which may be easily detected by the Huntsman Pathfinder system. The panel on the left shows a scenario for a rotating solar panel, which keeps the chassis pointing toward the Earth in a nominal configuration and resulting in a mean variation of 0.74mag, $\sigma = 0.43$ mag is seen. These results indicate it may be possible to variation in satellite attitude from nominal in some configurations with the Huntsman Pathfinder

## Appendix: Other Bright Satellites

While not the focus of this work, we also observe a range of other satellites that we choose from the 'brightest 100' Celestrak catalog[24] simply as targets of opportunity. A selection of the most complete passes and interesting objects are presented in Fig. 16. Target magnitude is presented as a function of time for easy of readability for these more complex passes, with solar phase angle as the colour bar.

The International Space Station is the only object for which we observe across the meridian flip, although flip time significantly impacts the completeness of the light curve. This transit path passes only 10 ° from the Sun at midday, illustrating the excellent stray light mitigation of the Canon lens and baffle hood.

Notably, Saocom 1A, the Atlas Centaur rocket body and the Chinese Space Station (Tiangong) all show signs of non-regular flashes, or momentary increases in brightness by $\sim 1 - 3$ mag on the order of milliseconds. Unstable rotation periods have been documented for the Atlas Centaur #694 by the Belgian Working Group for Satellites [42] which indicates that the space debris may be tumbling, but changing in rotation rate. It is unclear if these flashes may give insight into rotation of the body, or if they are similar to the Chinese Space Station and Saocom 1A which are more likely to be specular reflections off different parts of the chassis and solar panel arrays. Future work will include building on these observations to determine the total number of LEO satellites visible to the Huntsman Pathfinder system during the day.

---

[24] https://celestrak.org/

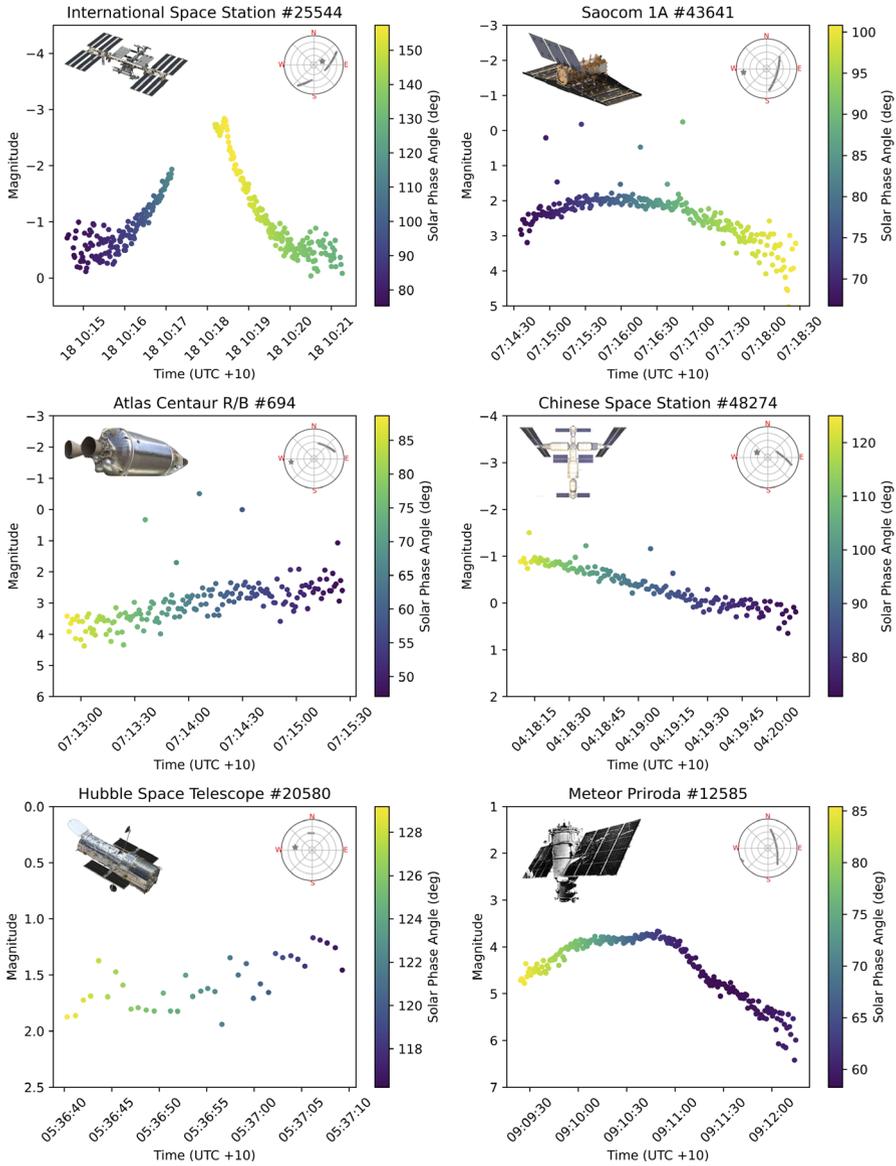

**Fig. 16** Target r-band magnitude as a function of time for a selection of non-Starlink objects with solar phase angle plotted as the colour bar. The transit path and sun location is showing in the transit diagram in the top right. As these targets are not the focus of this work, the transits are simply chosen at random as a target of opportunity from the Celestrak brightest 100 list. Image Credits—ISS: Konstantin Shaklein, SAOCOM 1A: ESA, Atlas Centaur: Chris Peat, Chinese Space Station: New York Times, Hubble Space Station: NASA, Meteor Priroda: Astronautix

**Acknowledgements** The authors would like to thank and acknowledged the assistance of Forrest Fankhauser, for his continued support of this work, and for providing insights and direction in the use of *lumos-sat*. We would like to thank Canon Australia for their support of the Huntsman Telescope collaboration and Pathfinder Huntsman project. We would also like to thank the team at the NASA's Langley Research Center for providing updates to the online CERES data ordering tool that enabled us to access 2024 CERES data products under short notice. The authors would like to acknowledge the generous assistance of Macquarie Observatory manager Adam Joyce who has provided continued technical support of this project that has made these results possible. We would also like to thank astro-imaging guru Jack Gow from Bintel for his continued support of this work and of the Huntsman Telescope project. We would also like to acknowledge the ever present, enthusiastic support of friends from the Association For Astronomy (AFA), Macquarie University. Finally the authors would like to acknowledge the traditional owners of the land on which the Huntsman Telescope is situated, Gamilarray, Wiradjuri and Wayilwan Country. In addition we would like to acknowledged the traditional owners of the land on which the Macquarie University Observatory is located, the Wallumattagal Clan of the Dharug Nation - whose cultures and customs have nurtured, and continue to nurture, this land since time immemorial.

**Author Contributions** Sarah E. Caddy: Conceptualization, Methodology, Software, Formal analysis, Investigation, Writing. Lee R. Spitler: Supervision, Conceptualization, Review & Editing.

**Funding** Open Access funding enabled and organized by CAUL and its Member Institutions. S.C. and L.S. acknowledge support from an Australian Research Council Discovery Project grant DP190102448

**Data Availability** Data collected with the Huntsman Pathfinder for the purposes of this work, and the python code used to reduce the data may be made available at the request of the first author.

## Declarations

**Conflict of interests** The authors report no conflict of interest to the best of our knowledge at the time of publication.

**Software** Lumos-sat (Fankhauser, F. et al. 2023) Astropy (The Astropy Collaboration 2013, 2018), Scipy (Virtanen, P. et al, 2020), Photutils (Bradley, L. et al, 2016), Source Extractor (Bertin, E. et al, 1996), Skyfield (Rhodes, B. et al, 2019),

# References

1. Australian Department of Defence: 2020 Force Structure Plan. Commonwealth Government of Australia Department of Defence, Canberra (2020)
2. United Nations Office for Outer Space Affairs: Guidelines for the Long-term Sustainability of Outer Space Activities of the Committee on the Peaceful Uses of Outer Space. United Nations Vienna Austria (2021). https://doi.org/10.18356/9789210021852
3. United States Space Force: Space Domain Awareness: Doctrine for Space Forces. United States Space Force. https://www.starcom.spaceforce.mil/Portals/2/SDP 3-100 Space Domain Awareness (November 2023)_pdf_safe.pdf (2023).
4. Lal, B., Balakrishnan, A., Caldwell, B., Buenconsejo, R., Carioscia, S. Global Trends in Space Situational Awareness and Space Traffic Management. Sci. Technol. Policy Inst. (2018)
5. Weeden, B., Cefola, P., Sankaran, J. Global space situational awareness sensors. In: Proceedings of the Advanced Maui Optical and Space Surveillance Technologies Conference. (2001)
6. Zimmer, P., Ackermann, M., McGraw, J. Optimizing Daylight Performance of Small Visible-NIR Optical Systems. In: Advanced Maui Optical and Space Surveillance Technologies Conference, Maui, Hawaii, USA (2020)
7. Estell, N., Ma, D., Seitzer, P.: Daylight Imaging of LEO Satellites Using COTS Hardware. In: Advanced Maui Optical and Space Surveillance Technologies Conference (2019)
8. Roggemann, M., Douglas, D., Therkildsen, E., Archambeault, D., Maeda, R., Schultz, D., Wheeler, B. (2010) Daytime Image Measurement and Reconstruction for Space Situational Awareness Applications. Advanced Maui Optical and Space Surveillance Technologies Conference, Wailea, Maui, Hawaii. pp. 17
9. Zimmer, P., McGraw, J., Ackermann, M.: Overcoming the Challenges of Daylight Optical Tracking of LEOs. Advanced Maui Optical and Space Surveillance Technologies Conference, Maui, Hawaii, USA (2021)
10. Gasparini, G., Miranda, V.: Space situational awareness an overview. In: Rathgeber, W., Schrogl, K.-U., Williamson, R.A. (eds.) The Fair and Responsible Use of Space An International Perspective Studies in Space Policy, pp. 73–87. Springer, Vienna (2010)
11. Qashoa, R., Lee, R.: Classification of low earth orbit (LEO) resident space objects' (RSO) light curves using a support vector machine (SVM) and long short-term memory (LSTM). sensors (Basel, Switzerland) 23(14), 6539 (2023). https://doi.org/10.3390/s23146539
12. Mariani, L., Cimino, L., Rossetti, M., Bucciarelli, M., Hossein, S.H., Varanese, S., Zarcone, G., Castronuovo, M., Di Cecco, A., Marzioli, P., Piergentili, F.: A dual perspective on geostationary satellite monitoring using DSLR RGB and sCMOS sloan filters. aerospace 10(12), 1026 (2023). https://doi.org/10.3390/aerospace10121026
13. Blacketer, L.D.J.: Attitude Characterisation of Space Objects using Optical Light Curves. PhD thesis, University of Southamp (2022)
14. Piergentili, F., Zarcone, G., Parisi, L., Mariani, L., Hossein, S.H., Santoni, F.: LEO object's light-curve acquisition system and their inversion for attitude reconstruction. Aerospace 8(1), 4 (2021). https://doi.org/10.3390/aerospace8010004
15. Kerr, E., Falco, G., Maric, N., Petit, D., Talon, P., Petersen, E.G., Dorn, C., Eves, S., Sánchez-Ortiz, N., Gonzalez, R.D., Nomen-Torres, J.: Light curves for geo object characterization. ESA Space Debris Office, Darmstadt, Germany (2021)
16. Koshkin, N., Shakun, L., Kozhukhov, O., Kozhukhov, D., Mamarev, V., Prysiaznyi, V., Ozeryan, A., Kudak, V., Neubauer, I.: Simultaneous multi-site photometry or leo satellites for rotation characterization. ESA Space Debris Office, Darmstadt, Germany (2021)
17. Jahirabadkar, S., Narsay, P., Pharande, S., Deshpande, G., Kitture, A. (2020). Space objects classification techniques A survey. In: 2020 International Conference on Computational Performance Evaluation (ComPE), pp. 786–791. https://doi.org/10.1109/ComPE49325.2020.9199996
18. Chote, P., Blake, J.A., Pollacco, D. Precision Optical Light Curves of LEO and GEO Objects. In: Advanced Maui Optical and Space Surveillance Technologies Conference (2019)
19. Skuljan, J. Photometric measurements of geostationary satellites over the Western Pacific Region. Advanced Maui Optical and Space Surveillance Technologies Conference, Maui, Hawaii, USA, (2018)

20. Silha, J., Pittet, J.-N., Hamara, M., Schildknecht, T.: Apparent rotation properties of space debris extracted from photometric measurements. Adv. Space Res. **61**(3), 844–861 (2018). https://doi.org/10.1016/j.asr.2017.10.048

21. Silha, J., Schildknecht, T., Pittet, J., Bodenmann, D., Kanzler, R., Karrang, P., Krag, H. Comparison of ENVISAT's Attitude Simulation and Real Optical and SLR Observations in order to Refine the Satellite Attitude Model. In: Proceedings of the Advanced Maui Optical and Space Surveillance Technologies Conference. (2016)

22. Binz, C.R., Davis, M.A., Kelm, B.E., Moore, C.I. Optical Survey of the Tumble Rates of Retired GEO Satellites. In: Advanced Maui Optical and Space Surveillance Technologies Conference (2014)

23. Papushev, P., Karavaev, Y., Mishina, M.: Investigations of the evolution of optical characteristics and dynamics of proper rotation of uncontrolled geostationary artificial satellites. Adv. Space Res. **43**(9), 1416–1422 (2009). https://doi.org/10.1016/j.asr.2009.02.007

24. Fankhauser, F., Tyson, J.A., Askari, J.: Satellite Optical Brightness. Astron. J. **166**(2), 59 (2023). https://doi.org/10.3847/1538-3881/ace047.

25. Matsushita, Y., Arakawa, R., Yoshimura, Y., Hanada, T.: Light Curve Analysis and Attitude Estimation of Space Objects Focusing on Glint. In: First Int'l. Orbital Debris Conf (2019)

26. Albuja, A.A., Scheeres, D.J., McMahon, J.W.: Evolution of angular velocity for defunct satellites as a result of YORP: An initial study. Adv. Space Res. **56**(2), 237–251 (2015). https://doi.org/10.1016/j.asr.2015.04.013

27. Caddy, S.E., Spitler, L.R., Ellis, S.C.: Toward a Data-driven Model of the Sky from Low Earth Orbit as Observed by the Hubble Space Telescope. Astron. J. **164**(2), 52 (2022). https://doi.org/10.3847/1538-3881/ac76c2.

28. Caddy, S.E., Spitler, L.R., Ellis, S.C.: An optical daytime astronomy pathfinder for the huntsman telescope. PASA **164**(2), 52 (2024). https://doi.org/10.3847/1538-3881/ac76c2.

29. Shaddix, J., Key, C., Ferris, A., Herring, J., Singh, N., Brost, T., Aristo?, J. Daytime Optical Contributions Toward Timely Space Domain Awareness in Low Earth Orbit. Advanced Maui Optical and Space Surveillance Technologies Conference, Maui, Hawaii, USA. (2021)

30. Kaminski, K., Zolnowski, M., Kaminska, M.K., Kruzynski, M., Kruzynska, D., Wnuk, E. Optimizing visual daytime satellite observations. (2021)

31. Alarcon, M.R., Licandro, J., Serra-Ricart, M., Joven, E., Gaitan, V., Sousa, R.D.: Scientific CMOS sensors in astronomy: IMX455 and IMX411. Publ. Astron. Soc. Pac. **135**(1047), 055001 (2023). https://doi.org/10.1088/1538-3873/acd04a

32. Rork, E.W., Lin, S.S., Yakutis, A.J.: Ground-based electro-optical detection of artificial satellites in daylight from reflected sunlight. NASA STI/Recon Tech. Rep. N. **83**, 10098 (1982)

33. Shaddix, J., Brannum, J., Ferris, A., Hariri, A., Larson, A., Mancini, T., Aristo?, J. Daytime GEO Tracking with "Aquila": Approach and Results from a New Ground-Based SWIR Small Telescope System. Advanced Maui Optical and Space Surveillance Technologies Conference, Maui, Hawaii, USA, (2019)

34. Horton, A., Spitler, L., Gee, W., Longbottom, F., Alvarado-Montes, J., Bazkiaei, A., Caddy, S., Lee, S.: The Huntsman Telescope: lessons learned from an autonomous telescope from COTS components. In: Proc. SPIE 11203, Advances in Optical Astronomical Instrumentation 2019, p 1120306 (2020) https://doi.org/10.1117/12.2539579

35. Rhodes, B. Skyfield: High precision research-grade positions for planets and Earth satellites generator. Astrophysics Source Code Library, ascl:1907.024. (2019).

36. The Astropy Collaboration, Price-Whelan, A.M., Lim, P.L., Earl, N., Starkman, N., Bradley, L., Shupe, D.L., Patil, A.A., Corrales, L., Brasseur, C.E., Nöthe, M., Donath, A., Tollerud, E., Morris, B.M., Ginsburg, A., Vaher, E., Weaver, B.A., Tocknell, J., Jamieson, W., Kerkwijk, M.H.v., Robitaille, T.P., Merry, B., Bachetti, M., Günther, H.M., Authors, P., Aldcroft, T.L., Alvarado-Montes, J.A., Archibald, A.M., Bódi, A., Bapat, S., Barentsen, G., Bazán, J., Biswas, M., Boquien, M., Burke, D.J., Cara, D., Cara, M., Conroy, K.E., Conseil, S., Craig, M.W., Cross, R.M., Cruz, K.L., D'Eugenio, F., Dencheva, N., Devillepoix, H.A.R., Dietrich, J.P., Eigenbrot, A.D., Erben, T., Ferreira, L., Foreman-Mackey, D., Fox, R., Freij, N., Garg, S., Geda, R., Glattly, L., Gondhalekar, Y., Gordon, K.D., Grant, D., Greenfield, P., Groener, A.M., Guest, S., Gurovich, S., Handberg, R., Hart, A., Hatfield-Dodds, Z., Homeier, D., Hosseinzadeh, G., Jenness, T., Jones, C.K., Joseph, P., Kalmbach, J.B., Karamehmetoglu, E., Kaluszynski, M., Kelley, M.S.P., Kern, N., Kerzendorf, W.E., Koch, E.W., Kulumani, S., Lee, A., Ly, C., Ma, Z., MacBride, C., Maljaars, J.M., Muna, D., Murphy, N.A., Norman, H., O'Steen, R., Oman, K.A., Pacifici, C., Pascual, S., Pascual-Granado,

J., Patil, R.R., Perren, G.I., Pickering, T.E., Rastogi, T., Roulston, B.R., Ryan, D.F., Rykoff, E.S., Sabater, J., Sakurikar, P., Salgado, J., Sanghi, A., Saunders, N., Savchenko, V., Schwardt, L., Seifert-Eckert, M., Shih, A.Y., Jain, A.S., Shukla, G., Sick, J., Simpson, C., Singanamalla, S., Singer, L.P., Singhal, J., Sinha, M., Sipocz, B.M., Spitler, L.R., Stansby, D., Streicher, O., Šumak, J., Swinbank, J.D., Taranu, D.S., Tewary, N., Tremblay, G.R., Val-Borro, M.d., Kooten, S.J.V., Vasovic, Z., Verma, S., Cardoso, J.V.d.M., Williams, P.K.G., Wilson, T.J., Winkel, B., Wood-Vasey, W.M., Xue, R., Yoachim, P., Zhang, C., Zonca, A., Astropy Project Contributors. The Astropy Project: Sustaining and Growing a Community-oriented Open-source Project and the Latest Major Release (v5.0) of the Core Package. Astrophys. J. **935**(2), 167 (2022). https://doi.org/10.3847/1538-4357/ac7c74

37. Greynolds, A.W. General physically-realistic BRDF models for computing stray light from arbitrary isotropic surfaces. Optical Modeling and Performance Predictions VII, 9577: 97–104. (2015). https://doi.org/10.1117/12.2185080

38. Stackhouse, P., Gupta, S., Kratz, D., Geier, E., Edwards, A., Wilber, A. Fast Longwave and Shortwave Radiative Fluxes (FLASHFlux) From CERES and MODIS Measurements. 37th COSPAR Scientific Assembly 3014 (2008)

39. Horiuchi, T., Hanayama, H., Ohishi, M., Nakaoka, T., Imazawa, R., Kawabata, K.S., Takahashi, J., Onozato, H., Saito, T., Yamanaka, M., Nogami, D., Tampo, Y., Kojiguchi, N., Ito, J., Shibata, M., Schramm, M., Oasa, Y., Kanai, T., Oide, K., Murata, K.L., Hosokawa, R., Takamatsu, Y., Imai, Y., Ito, N., Niwano, M., Takagi, S., Ono, T., Kouprianov, V.V.: Multicolor and multi-spot observations of Starlink's Visorsat. Publ. Astron. Soc. Japan **75**(3), 584–606 (2023). https://doi.org/10.1093/pasj/psad021

40. Goldman, D.: IBFS File Nos. SAT-LOA-20200526-00055 and SAT-AMD-20210818-00105. https://forum.nasaspaceflight.com/index.php?action=dlattach;topic=46726.0;attach=2103197 (2022).

41. DalBello, R.: Re: SpaceX Comment-Office of Space Commerce, National Oceanic and Atmospheric Administration, Department of Commerce, Request for Information (RFI) on Scope of Civil Space Situational Awareness Services (88 FR 4970) https://www.space.commerce.gov/wp-content/uploads/2023-02-27-SpaceX.pdf (2023).

42. Pontieu, B.: Database of photometric periods of artificial satellites. Adv. Space Res. **19**, 229–232 (1997). https://doi.org/10.1016/S0273-1177(97)00005-7

## Authors and Affiliations

**Sarah E. Caddy[1,2,3]**  **· Lee R. Spitler[3,4]**

✉ Sarah E. Caddy
  sarah.caddy@unimelb.edu.au

1   Melbourne Space Laboratory, University of Melbourne, Melbourne, VIC 3010, Australia

2   School of Mathematical and Physical Sciences, Macquarie University, Sydney, NSW 2109, Australia

3   Australian Astronomical Optics, Macquarie University, Sydney, NSW 2113, Australia

4   Astrophysics and Space Technologies Research Centre, Macquarie University, Sydney, NSW 2109, Australia